\documentclass[12pt]{article}
\pdfoutput=1
\usepackage[utf8]{inputenc}

\usepackage{xcolor}
\usepackage{amsmath,amssymb,mathrsfs,bm,setspace,xspace,soul,empheq}  
\usepackage{graphicx}
\usepackage{physics}  
\usepackage{colonequals} 
\usepackage{float}  
\usepackage{comment}     
\usepackage{siunitx}       
\usepackage{tablefootnote}

\usepackage{subcaption}

\usepackage{braket}
\usepackage{color}  
\usepackage{dcolumn}
\usepackage{multirow}            
\usepackage{geometry}         
\usepackage{tabularx}
\definecolor{darkblue}{rgb}{0.1,0.1,.7}
\usepackage[colorlinks, linkcolor=darkblue, citecolor=darkblue, urlcolor=darkblue, linktocpage,backref=page]{hyperref} 
\renewcommand*{\backref}[1]{}
\renewcommand*{\backrefalt}[4]{%
	\ifcase #1 (Not cited.)%
	\or        (Cited on p.~#2.)%
	\else      (Cited on pp.~#2.)%
	\fi}

\usepackage{amsmath,amssymb,graphicx,enumerate,bbm}
\usepackage{booktabs}
\usepackage{geometry}
\usepackage[T1]{fontenc} 

\usepackage{tikz}
\usetikzlibrary{spy}

\usepackage{newfloat}
\DeclareFloatingEnvironment[
    listname={Algorithm},
    name=Algorithm,
    placement=tbhp,
]{algorithmfloat}

\usepackage[boxed]{algorithm2e}

\definecolor{martenorange}{rgb}{0.91, 0.41, 0.17}

\usepackage[english]{babel}
\usepackage{csquotes}
\MakeOuterQuote{"}

\makeatletter
\newcommand{\subalign}[1]{%
  \vcenter{%
    \Let@ \restore@math@cr \default@tag
    \baselineskip\fontdimen10 \scriptfont\tw@
    \advance\baselineskip\fontdimen12 \scriptfont\tw@
    \lineskip\thr@@\fontdimen8 \scriptfont\thr@@
    \lineskiplimit\lineskip
    \ialign{\hfil$\m@th\scriptstyle##$&$\m@th\scriptstyle{}##$\hfil\crcr
      #1\crcr
    }%
  }%
}
\makeatother
\usepackage[shortlabels,inline]{enumitem}

\newcommand\x{\times}

\newcommand{\N}{\calN}

\newcommand{\beq}{\begin{equation}} 
\newcommand{\eeq}{\end{equation}}
 
\def\nn{\nonumber}

\def\bZ {\mathbb{Z}}

\def\cO{{\cal O}}

\def\calN {{\cal N}} 
\def\cN {{\cal N}}

\def\cP {{\cal P}}

\def\ge{\geqslant}

\def\geq{\geqslant}
\def\leq{\leqslant}
\def\<{\langle}
\def\>{\rangle}
\def\cO{{\cal O}}

\def\De{\Delta}

\def\e{\epsilon}
\def\s{\sigma}

\numberwithin{equation}{section}
\begin{document}
\vspace*{-.6in} \thispagestyle{empty}
\vspace{1cm} 
{\Large
	\begin{center}
		{\bf Rigorous bounds on irrelevant operators in the 3d Ising model CFT}
	\end{center}
}
\vspace{1cm}
\begin{center}
	{\bf Marten Reehorst$^{a,b}$
	}\\[2cm] 
	{
		$^a$  Institut des Hautes \'Etudes Scientifiques, 91440 Bures-sur-Yvette, France\\
		$^b$ CPHT, CNRS, \'Ecole Polytechnique, Institut Polytechnique de Paris,\\Route de Saclay, 91128 Palaiseau, France
	}
	\vspace{1cm}
\end{center}

\vspace{4mm}
\begin{abstract}
We use the recently developed navigator method to obtain rigorous upper and lower bounds on new OPE data in the 3d Ising CFT. For example, assuming that there are only two $\mathbb{Z}_2$-even scalar operators $\epsilon$ and $\epsilon'$ with a dimension below 6 we find a narrow allowed interval for $\Delta_{\epsilon'}$, $\lambda_{\sigma\sigma\epsilon'}$ and $\lambda_{\epsilon\epsilon\epsilon'}$. With similar assumptions in the $\mathbb{Z}_2$-even spin-2 and the $\mathbb{Z}_2$-odd scalar sectors we are also able to constrain: the central charge $c_T$; the OPE data $\Delta_{T'}$, $\lambda_{\e\e T'}$ and $\lambda_{\s\s T'}$ of the second spin-2 operator; and the OPE data $\Delta_{\sigma'}$ and $\lambda_{\sigma\epsilon\sigma'}$ of the second $\mathbb{Z}_2$-odd scalar. We compare the rigorous bounds we find with estimates that have been previously obtained using the extremal functional method (EFM) and find a good match. This both validates the EFM and shows the navigator-search method to be a feasible and more rigorous alternative for estimating a large part of the low-dimensional operator spectrum. We also investigate the effect of imposing sparseness conditions on all sectors at once. We find that the island does not greatly reduce in size under these assumptions. We efficiently find islands and determine their size in high-dimensional parameter spaces (up to 13 parameters). This shows that using the navigator method the numerical conformal bootstrap is no longer constrained to the exploration of small parameter spaces.
\end{abstract}


\vspace{.2in}
\vspace{.3in}
\hspace{0.2cm} 

\newpage

{\small
	\parskip=-0.1em
\tableofcontents
}

\newpage
\section{Introduction}
\label{frombootstrap}
Recently a new "navigator" method was developed for the numerical conformal bootstrap and was shown to be efficient at finding an allowed point in the 3d ising model island as well as finding the island's boundaries \cite{Reehorst:2021ykw}. The navigator method replaces the boolean "excluded"/"allowed" test of an assumed CFT spectrum with a "navigator" function that gives a continuous measure of success. A spectrum assumption can be parameterized by a set of numbers $\bf{p}$ in a search space $\cP$. This search space must always include the external dimensions but can also parametrize other assumption such as a gap that is imposed in a certain sector or the dimensions or OPE coefficients of isolated operators. The navigator $\N(\mathbf{p})$ gives positive values for excluded points $\bf{p}$ and negative values for allowed points. Moreover, it is bounded from above by 1 by construction.\footnote{In this paper we use the GFF construction of the navigator in which the upper bound can be chosen to be one. When using the $\Sigma$ navigator instead the navigator is still bounded but the exact bounding value is not known a priori.} By inspection the example 2- and 3-parameter navigator functions studied in \cite{Reehorst:2021ykw} 
were also shown to be $C^1$ differentiable and to only possess a single minimum in the vicinity of the Ising model. It was also shown that the gradient $\nabla \N(\bf{p})$ can cheaply be computed wherever $\N(\bf{p})$ has been computed. 
In order to find an allowed point the navigator $\N(\bf{p})$ was minimized locally, using a modified BFGS algorithm. This algorithm makes use of the gradient information at various points to reconstruct a quadratic model which it minimizes at every step. In order to deal with inaccuracies of the quadratic model a line search is used to ensure sufficient decrease in each step as well as positive definiteness of all updates to the Hessian in the quadratic model. In this work we apply all these new tools to the task of obtaining new rigorous bounds on parts of the Ising model spectrum for which previously only non-rigorous estimates were known \cite{El-Showk:2014dwa,Komargodski:2016auf,Simmons-Duffin:2016wlq}.

A major bottleneck in the numerical conformal bootstrap using conventional scanning methods was that both the problem of finding an allowed point and the problem of finding the boundaries of the allowed "island" around that point scale badly with the dimensionality $N$ of the search space. In \cite{Reehorst:2019pzi} it was hypothesized that navigator search methods would scale much better with the dimensionality of the search space, but it was only applied to low dimensional search spaces with $N=2,3$. In this paper we apply navigator based search algorithms to problems involving much larger parameter spaces and show that the problem scales much better with $N$ than traditional scanning methods. The navigator methods thus offers a major breakthrough enabling us to tackle more difficult bootstrap problems involving more parameters. In this work we successfully obtain islands in search spaces of dimensions $4$ to $13$, leading to the rigorous upper and lower bounds on the dimensions and OPE coefficients of the operators $\epsilon'$, $\sigma'$ and $T'$, see Table ~\ref{table:summary_all_strongest_bounds}.\footnote{Previously, $\Delta_{\epsilon'}$, $\Delta_{\sigma'}$,$\lambda_{\sigma\sigma\epsilon'}$,$\lambda_{\epsilon\epsilon\epsilon'}$,$\lambda_{\sigma\epsilon\sigma'}$, $\Delta_{T'}$, $\lambda_{\epsilon\epsilon T'}$ and $\lambda_{\sigma\epsilon T'}$ were only known from non-rigorous estimates obtained by the EFM \cite{Simmons-Duffin:2016wlq}, while for $c_T$ only a rigorous upper bound was known \cite{El-Showk:2014dwa}.}

\begin{table*}[h!]
	\centering
	\begin{tabular}{@{}l l l@{}}
		\toprule
		& Rigorous bounds \\
		\midrule
		$\Delta_\sigma$ & $0.518157(\mathbf{35})$ & $\Lambda=19$ \\
	    $\Delta_\epsilon$ & $1.41265(\mathbf{36}) $ & $\Lambda=19$ \\
		$\lambda_{\sigma\sigma\epsilon}$ & $1.05185(\mathbf{12})$ & $\Lambda=19$\\
		$\lambda_{\epsilon\epsilon\epsilon}$ & $1.53240(\mathbf{58})$ & $\Lambda=19$ \\
		$\Delta_{\epsilon'}$& $3.82951(\mathbf{61})$ & $\Lambda=31 $ \\
		$\lambda_{\sigma\sigma\epsilon'}$ &$0.05304(\mathbf{16})$ &  $\Lambda=19$ \\
		$\lambda_{\epsilon\epsilon\epsilon'}$ & $1.5362(\mathbf{12})$ &  $\Lambda=19$ \\
		$\Delta_{\sigma'}$& $5.262(\mathbf{89})$ & $\Lambda=19 $ \\
		$\lambda_{\sigma\epsilon\sigma'}$&$0.0565(\mathbf{15})$ & $\Lambda=19 $ \\
		$\frac{c_T}{c_T^{\textrm{free}}}$ & $0.946543(\mathbf{42})$  & $\Lambda=19 $ \\
        $\Delta_{T'} $ & $5.499(\mathbf{17})$  & $\Lambda=19 $ \\
        $\lambda_{\s\s T'}$ & $0.02107(\mathbf{20})$ & $\Lambda=19$ \\
        $ \lambda_{\e\e T'}$ & $1.355(\mathbf{30})$ & $\Lambda=19$ \\
		\bottomrule 
	\end{tabular}
	\caption{A summary of the strongest rigorous bounds found in this paper using various sparseness assumptions. All errors (presented in bold) are rigorous. The bounds on the first four quantities were already known \cite{Kos:2016ysd}. The other rigorous bounds are new. 
	}
	\label{table:summary_all_strongest_bounds}
\end{table*}
Heretofore, the extremal functional method (EFM) was the main method for obtaining a large amount of information on the entire low dimensional spectrum \cite{ElShowk:2012hu,El-Showk:2014dwa,Komargodski:2016auf,Simmons-Duffin:2016wlq}. Given a sufficiently small island the zeros of an extremal functional are believed to give good estimates for the dimensions of many low dimensional exchanged operators. However, these estimates come without any rigorous error bars.  

This work shows that the navigator function is an efficient alternative for obtaining bounds on a large number of parameters. Moreover, the bounds found using the navigator method are rigorous, unlike those obtained through the EFM. 

At the same time this work provides an essential test of the EFM. For all quantities studied in this work we find only one isolated allowed interval that matches the results previously obtained in \cite{Simmons-Duffin:2016wlq}. Not only do we find bounds consistent with the EFM estimates but the rigorously obtained allowed intervals also seem to be very similar to the error bars estimated using the EFM method. 

Although our methods are completely general, in this work we are exclusively concerned with the OPE data of the three dimensional Ising model. As usual, we denote the lowest dimensional $\bZ_2$-odd and $\bZ_2$-even scalars as $\sigma$ and $\epsilon$ respectively and the stress tensor as $T$. We also denote the next lowest dimensional operators of the same type by adding a prime, i.e $\epsilon'$, $\sigma'$ and $T'$, and so forth for the next to next to lowest dimensional operators $\epsilon''$, $\sigma''$ and $T''$.

This paper is structured as follows: In Section~\ref{sec:bootstrap_setup} we discuss the bootstrap setup we used (with Appendix~\ref{app:navigators} containing further details on the exact construction of all the different navigator functions and Appendix~\ref{app:parameters_numerics} containing further details on the choice of numerical parameters). In Section~\ref{sec:method} we review the navigator search methods and explain how we applied them to searches of high-dimensional parameter spaces. In Section~\ref{sec:results} we discuss the results we obtained using these methods. We conclude in Section~\ref{sec:conclusions}.

\section{Bootstrap setup}
\label{sec:bootstrap_setup}

We start from the standard mixed bootstrap of the $\langle\sigma \sigma\sigma\sigma\rangle$, $\langle\sigma \sigma\epsilon\epsilon\rangle$ and $\langle\epsilon \epsilon\epsilon\epsilon\rangle$ correlations functions where $\sigma$ and $\epsilon$ are respectively $\mathbb{Z}_2$-odd and $\mathbb{Z}_2$-even scalars in a $\mathbb{Z}_2$-invariant CFT. This setup was previously used among others to find rigorous high precision bounds on $\Delta_{\sigma}$ and $\Delta_{\epsilon}$ \cite{ElShowk:2012ht,El-Showk:2014dwa,Kos:2014bka,Simmons-Duffin:2015qma,Kos:2016ysd} and to obtain estimates of the dimensions and OPE coefficients of many operator dimensions and OPE coefficients using the EFM and the light-cone bootstrap \cite{Simmons-Duffin:2016wlq}.

The "navigator-improved" bootstrap equations for such a $\mathbb{Z}_2$-invariant CFT can be written as
\begin{multline}
\label{eq:crossingIsingWithM}
   \vec{V}_{0,0} + \lambda \vec{M} +
   \text{Tr}
   \left[
   P_{\Delta_\e,0} \left(
    \vec{V}_{+,\Delta_\e,0}
    +\begin{pmatrix}1&0\\ 0 & 0\end{pmatrix}
    \vec{V}_{-,\Delta_\sigma,0}
    \right) 
    \right] \\
    +\sum_{(\Delta,\ell)\in S_+}  \Tr[P_{\Delta, \ell} \vec{V}_{+,\De,\ell}] + \sum_{(\Delta,\ell)\in S_-} p_{\Delta,\ell} \vec{V}_{-,\De,\ell} = 0\,,
\end{multline}
These crossing equations where first written, without the $\lambda \vec{M}$ term, in \cite{ElShowk:2012ht}. Here we instead follow the notation of \cite{Reehorst:2021ykw}. For the precise definition of $P_{\Delta_\e,0}$, $p_{\Delta,\ell}$, $\vec{V}_{0,0}$, $\vec{V}_{+,\Delta_\e,\ell}$ and  $\vec{V}_{-,\Delta_\sigma,\ell}$  see equations (2.14) to (2.17) of that work. $S_-$ and $S_+$ refer to some set of $(\Delta,\ell)$ allowed by our assumptions on respectively the $\bZ_2$-odd/even part of the CFT spectrum. We will specify them on a case by case basis for the various assumptions that we will study. The term $\lambda \vec{M}$ is traditionally absent and is added here, following \cite{Reehorst:2021ykw}, so that the above equation is guaranteed to have a solution for any value of the external operators $\Delta_\sigma$ and $\Delta_{\epsilon}$ as well as for any choice of $S_-$ and $S_+$, i.e. for any of the spectrum assumptions that we will be making. This guarantees that we can write an optimization task that outputs a navigator function that is bounded from above, namely 
\begin{equation}
    \N(\bf{p})=\min \lambda \quad \text{ such that equation \eqref{eq:crossingIsingWithM} has a solution}.
\end{equation} 
Here $\bf{p}$ is a vector that parameterizes the external dimensions and spectrum assumptions being tested. We will be using the GFF-construction of the navigator function introduced in section 2.1.1 of \cite{Reehorst:2021ykw}.

The exact form of $\vec{M}$ will depend on the number of GFF-operators not present in $S_-$ and $S_+$ and will thus depend on the specific question we are solving. The general procedures is to sum all GFF operators not present in $S_-$ and $S_+$ weighted by their OPE coefficients:
\begin{align}
    \label{eq:GFF-general}
          \vec{M}_{\rm GFF} = &\sum_{(n,\ell)^{\s\s}} 
        \text{Tr}\hspace{-3pt}\left[
        \begin{pmatrix} \lambda^2_{\sigma\sigma (2\Delta_\sigma+2n+\ell, \ell)} &0\\0&0 \end{pmatrix}     \vec{V}_{+,{2\Delta_\sigma+2n+\ell},\ell} \right]
        \hspace{-2pt}+\hspace{-1pt}\text{Tr}\hspace{-3pt}\left[\sum_{(n,\ell)^{\e\e}} \hspace{-2pt}
         \begin{pmatrix} 0 &0\\0& \lambda^2_{\epsilon\epsilon(2\Delta_\epsilon+2n+\ell, \ell)} \end{pmatrix} \hspace{-2pt} \vec{V}_{+,{2\Delta_\epsilon+2n+\ell},\ell}   \right] \nonumber \\
         +&\qquad \sum_{(n,\ell)^{\s\e}}  \lambda^2_{\sigma\epsilon (\Delta_\sigma+\Delta_\epsilon 2n+\ell,\ell)} \vec{V}_{-,\Delta_\s+\Delta_\e+2n+\ell,\ell}\,,\\
         &\qquad \text{with: } (n,\ell)^{\cO_1 \cO_2} = \begin{cases}
         \{n,\ell \in \mathbb{Z} : \Delta_{\cO_1} +\Delta_{\cO_2} +2n+\ell < \Delta_{n,\ell,+}^*\} & \text{ if } \cO_1=\cO_2\\
         \{n,\ell \in \mathbb{Z} : \Delta_{\cO_1} +\Delta_{\cO_2} +2n+\ell < \Delta_{n,\ell,-}^*\} & \text{ if } \cO_1\neq\cO_2
         \end{cases}
         \nonumber
\end{align}
where $\Delta_{n,\ell,\pm}^*$ is the assumed gap in the $\bZ_2$-even/odd spin $\ell$ sector. The OPE coefficients are taken from the known analytic expressions \cite{Fitzpatrick:2011dm,Fitzpatrick:2012yx,Karateev:2018oml}: 
\begin{align}
\label{eq:scalar_GFF_coefficients}
    &\lambda^2_{\sigma\sigma (2\Delta _{\sigma } +2n+\ell,\ell)} = \frac{(-2)^{\ell } \left((-1)^{\ell }+1\right) \left(\left(-\frac{d}{2}+\Delta _{\sigma }+1\right)_n \left(\Delta _{\sigma }\right)_{n+\ell } \right)^2 }{n! \ell ! \left(\frac{d}{2}+\ell \right)_n \left(-d+n+2 \Delta _{\sigma }+1\right)_n \left(2 n+\ell +2 \Delta _{\sigma }-1\right)_{\ell } \left(-\frac{d}{2}+n+\ell +2 \Delta _{\sigma }\right)_n}\\
    &\lambda^2_{\sigma\epsilon (\Delta _{\sigma }+ \Delta _{\epsilon }+ 2n+\ell,\ell)} = \frac{(-1)^{n}2^{\ell} \Gamma (\frac{d}{2}-\Delta_{\sigma}) \Gamma (\frac{d}{2}-\Delta_{\epsilon}) \Gamma (\frac{d}{2}+\ell) \Gamma (\ell+n+\Delta_{\sigma}) \Gamma (\ell+n+\Delta_{\epsilon}) 
}{
\Gamma (\Delta_{\sigma}) \Gamma (\Delta_{\epsilon}) \Gamma (\ell+1) \Gamma (n+1) \Gamma (\frac{d}{2}+\ell+n) \Gamma (\frac{d}{2}-n-\Delta_{\sigma}) \Gamma (\frac{d}{2}-n-\Delta_{\epsilon})}
 \nn\\
& \hspace{12pt} \x \frac{\Gamma (d-2 n-\Delta_{\sigma}-\Delta_{\epsilon}) \Gamma (\ell+2 n+\Delta_{\sigma}+\Delta_{\epsilon}-1) \Gamma (-\frac{d}{2}+\ell+n+\Delta_{\sigma}+\Delta_{\epsilon})
}{
 \Gamma (d-n-\Delta_{\sigma}-\Delta_{\epsilon}) \Gamma (2 \ell+2 n+\Delta_{\sigma}+\Delta_{\epsilon}-1) \Gamma (-\frac{d}{2}+J+2 n+\Delta_{\sigma}+\Delta_{\epsilon})
}
\end{align}
where $(a)_{n}$ denotes the Pochhammer symbol.\footnote{The factor of $(-2)^\ell$ difference compared to equation (11) of \cite{Fitzpatrick:2012yx} is because we use a different conformal block normalization. We use the same normalization as \cite{Poland:2018epd}, see equation (52) of that work for details.} Of course $\lambda^2_{\epsilon\epsilon (2\Delta _{\epsilon }+ 2n+\ell)}$ is the same as $\lambda^2_{\sigma\sigma (2\Delta _{\sigma }+2n+\ell)}$ but with $\Delta_{\sigma}$ replaced by $\Delta_{\epsilon}$.
\label{eq:GFFGeneral}

In addition we can impose a relationship between the OPE coefficients $\lambda_{\sigma \sigma \epsilon}$ and $\lambda_{\epsilon \epsilon \epsilon}$ of the form   $\lambda_{\epsilon \epsilon \epsilon} =\tan(\theta) \lambda_{\sigma \sigma \epsilon}$ for some value of an "OPE angle" $\theta$. In this case the third term in equation \eqref{eq:crossingIsingWithM} can be written as
\begin{equation}
    \begin{pmatrix}\cos(\theta) & \sin(\theta)\end{pmatrix} \cdot \left(
    \vec{V}_{+,\Delta_\e,0}
    +\begin{pmatrix} 1 & 0 \\ 0 & 0\end{pmatrix}
    \vec{V}_{-,\Delta_\sigma,0}
    \right)
    \cdot \begin{pmatrix}\cos(\theta)\\ \sin(\theta) \end{pmatrix}.
\end{equation}

Finally, we can also test the feasibility of OPE magnitudes, not just of OPE ratios. As an example let's take the OPE coefficients appearing in the exchange of the stress tensor $\lambda_{\sigma\sigma T}$ and $\lambda_{\epsilon\epsilon T}$. These are known to be related by the Ward identity $\frac{\lambda_{\sigma\sigma T}}{\Delta_\sigma}=\frac{\lambda_{\epsilon\epsilon T}}{\Delta_\epsilon}$ leaving only one unknown parameter to be determined by the numerical bootstrap. In order to be able to scan over this parameter we can contract the matrix $V_{+,3,2}$ twice with the vector $(\Delta_\sigma,\Delta_\epsilon)$ and call this $\tilde{V}_{+,3,2}$. Then we can consider the crossing equation
\begin{multline}
\label{eq:crossing_ct}
   \vec{V}_{0,0} +\tilde{p}_T \tilde{V}_{+,3,2} + \lambda \vec{M} +
   \text{Tr}
   \left[
   P_{\Delta_\e,0} \left(
    \vec{V}_{+,\Delta_\e,0}
    +\begin{pmatrix}1&0\\ 0 & 0\end{pmatrix}
    \vec{V}_{-,\Delta_\sigma,0}
    \right) 
    \right] \\
    +\sum_{(\Delta,\ell)\in S_+}  \Tr[P_{\Delta, \ell} \vec{V}_{+,\De,\ell}] + \sum_{(\Delta,\ell)\in S_-} p_{\Delta,\ell} \vec{V}_{-,\De,\ell} = 0\,,
\end{multline}
Here $\tilde{p}_T = \frac{\lambda^2_{\sigma\sigma T}}{\Delta_\sigma^2}$ will be a parameter that the navigator function will depend on. In order to be able to find lower bounds on $\tilde{p}_T$, in addition to an upper bound, it is important that $S_+$ has some gap above the stress tensor in the spin 2 sector. Note that this equation assumes the existence of a stress tensor with an OPE coefficients of exactly $\tilde{p}_T$. The GFF solution will not contain this operator with these OPE coefficients so a small tweak to $\vec{M}_{\rm GFF}$ is required in order for the $\lambda \vec{M}$ term to still guarantee the existence of a solution with $\lambda=1$ and thus guarantee boundedness of the navigator function. In general, for any assumed contribution $p_f V_f$ we can guarantee a solution to \eqref{eq:crossing_ct} with $\lambda=1$ by using the following choice of $\vec{M}$:
\begin{equation}
\label{eq:fixed_term}
    \vec{M}_{\rm{GFF},f}=\vec{M}_{\rm GFF} - p_f V_f
\end{equation}
In this example $p_f V_f=\tilde{p}_T \tilde{V}_{+,3,2}$.

The major difference between this work and previous studies of these bootstrap equations will be in the assumptions on the allowed CFT spectrum described by $S_{+}$ and $S_{-}$. Most notably the navigator function method enables us to consider spectrum assumptions that depend on a large number of parameters. This enables us to scan and thus bound additional parameters such as the dimensions and OPE coefficients of $\e'$, $\s'$ and $T'$.
 
However, in order to be able to find both upper and lower bounds on the dimension of these irrelevant operators (and to be able to bound their OPE coefficients) we have to make an assumption that isolate these operators from the rest of the spectrum. In this work we do this by assuming there are only two operators with a dimension below $6$ in the sectors we study.

The choice of a gap of $6$ that is used throughout this paper was made because it is a nice round number such that the Ising model is expected to contain exactly 2 operators below it in the three sectors that we will be studying in the subsections of section \ref{sec:results} 
\begin{equation*}
    \textrm{\ref{sec:results:even-scalar})  $\bZ_2$-even scalars   $\quad$ \ref{sec:results:odd-scalar}) $\bZ_2$-odd scalars $\quad$  \ref{sec:results:even-spin2}) $\bZ_2$-even spin-2 operators}
\end{equation*}
This expectation is mainly based on the spectrum obtained in \cite{Simmons-Duffin:2016wlq} via a non-rigorous extremal functional method where $\Delta_{\e''} \approx 6.8956(43)$, $\Delta_{T''}\approx 7.0758(58)$ and $\Delta_{\sigma''}>8$.\footnote{Besides $\s$ and $\s'$ no additional (stable) zeros corresponding to $\bZ_2$-odd scalar were found below 8.} Since this estimate was obtained using a non-rigorous method it could be wrong. However, for the number of operators with a dimension below 6 to stay the same, the estimates of the spectrum obtained by EFM method only need to be somewhat accurate. 

We can also compare this assumption against the free theory spectrum and perturbative results from a $4-\epsilon$ expansion. Table \ref{table:epsilonExp} shows that these assumptions hold for the free theory as well as in a naive first order epsilon expansion.\footnote{Of course the $\epsilon$-expansion is not expected to be very accurate at $\e=1$. However, it might still give a somewhat accurate account of the number of operators in each sector with a dimension below 6. We only include the first order result since without a proper re-summation the higher order terms do not give sensible results when evaluated at  $\epsilon=1$.}

\begin{table*}[h!]
	\centering
	\begin{tabular}{@{}l | l l l l @{}}
		\toprule
  & $ \epsilon$ -expansion  & $ \epsilon =0 $ & $ \epsilon =1 $\\
   \midrule
$ \Delta_{\sigma } $ & $ 1-\frac{\epsilon }{2}+O\left(\epsilon ^2\right) $ & $ 1 $ & $ 0.5 $\\
$ \Delta_{\sigma '} $ & $ 5+\frac{5 \epsilon }{6}+O\left(\epsilon ^2\right) $ & $ 5 $ & $ 5.83333 $\\
$ \Delta_{\sigma ''} $ & $ 7+\frac{7 \epsilon }{2}+O\left(\epsilon ^2\right) $ & $ 7 $ & $ 10.5 $\\
$ \Delta_{\sigma '''} $ & $ 9-\frac{5 \epsilon }{18}+O\left(\epsilon ^2\right) $ & $ 9 $ & $
8.72222 $\\
 \midrule
$ \Delta_{\epsilon } $ & $ 2-\frac{2 \epsilon }{3}+O\left(\epsilon ^2\right) $ & $ 2 $ & $ 1.33333 $\\
$ \Delta_{\epsilon '} $ & $ 4 +O\left(\epsilon ^2\right)  $ & $ 4 $ & $ 4. $\\
$ \Delta_{\epsilon ''} $ & $ 6+2 \epsilon +O\left(\epsilon ^2\right) $ & $ 6 $ & $ 8. $\\
$ \Delta_{\epsilon '''} $ & $ 8-\frac{8 \epsilon }{9}+O\left(\epsilon ^2\right) $ & $ 8 $ & $ 7.11111 $\\
 \midrule
$ \Delta_{T} $ & $ 4-\epsilon +O\left(\epsilon ^2\right) $ & $ 4 $ & $ 3. $\\
$ \Delta_{T'} $ & $ 6-\frac{5 \epsilon }{9}+O\left(\epsilon ^2\right) $ & $ 6 $ & $ 5.44444 $\\
$ \Delta_{T''} $ & $ 8-\frac{10 \epsilon }{9}+O\left(\epsilon ^2\right) $ & $ 8 $ & $ 6.88889 $\\
$ \Delta_{T'''} $ & $ 8+\frac{11 \epsilon }{9}+O\left(\epsilon ^2\right) $ & $ 8 $ & $ 9.22222 $\\
    \bottomrule
	\end{tabular}
	\caption{The dimensions of the lowest dimensional $\bZ_2$-odd and $\bZ_2$-even scalars and $\bZ_2$-even spin-2 operators in the 3d Ising model according to the $\e$-expansion (up to order $O(\epsilon)$) \cite{Kehrein:1992fn,Johan:upcoming}. The free theory (i.e. $\epsilon=0$) values are shown in the second column while the evaluation of the first order $\e$-expansion at $\epsilon=1$ is shown in the last column. While the latter estimates are of course not very accurate at $\epsilon=1$, they offer (weak) support that each of these sectors contains two operators with a dimension below 6.
	}
	\label{table:epsilonExp}
\end{table*}

We also test the dependence of the bounds on the exact value of the imposed gap on $\Delta_{\e''}$ in section \ref{sec:results:even-scalar} and find that the value of the gap only very slightly affects the bounds as long as the gap is sufficiently below the EFM estimate.\footnote{To gain even more confidence in the gap assumptions one could follow the dimensions of these operators in $4-\epsilon$ space-time dimensions to $d=3$ from some small value of epsilon that is well described by perturbation theory and where one can thus be completely confident in the chosen gap assumption.}

In addition to these additional sparseness assumptions in these three sectors we will also always assume that there is only one relevant $\bZ_2$-even and one $\bZ_2$-odd scalar as this has been firmly established by experimental evidence.

Finally we study the effect of making all these assumptions in the different sectors at the same time. One could hope that imposing sparseness in all these sectors simultaneously could be highly constraining and lead to a great reduction in the size of the Ising island. We investigate this and find it not to be the case: Imposing the sparseness assumptions in all these sectors at once does not lead to a major reduction in the size of any of the bounds. This was somewhat to be expected. Apart from the presence of some spurious poles, the Ising spectrum as extracted by the EFM already looks sparse for all allowed points at least when extracted at sufficiently high values of the derivative order $\Lambda$. Imposing sparseness assumptions in multiple sectors could still be useful to isolate theories that are not isolated as easily by the numerical conformal bootstrap using only more basic assumptions.

Throughout this paper we study many navigator functions using different spectrum assumptions. To keep the notation simple we will simply denote a certain navigator function by the parameters that it depends on. When an operator dimension, besides the external ones, appears as an argument of the navigator function this indicates we are assuming that it is isolated from other similar operators by a gap of 6. For example $\N(\Delta_\epsilon,\Delta_\sigma,\theta,\Delta_{\epsilon'})$, corresponds to a bootstrap problem where we assume a gap $\Delta_{\epsilon''}>6$ (as well as $\Delta_{\sigma'}>3$ which we always assume). All details of the various navigator functions we studied can be found in Appendix~\ref{app:navigators}.

\section{Method}
\label{sec:method}
Similar to traditional feasibility based methods we first want to identify isolated allowed regions and then find the borders of these regions. In order to do this we minimize the navigator from various different initial points in order to (hopefully) identify all the isolated allowed (i.e. negative) regions. 

In fact, the expectation is that for a $\bZ_2$-invariant CFT and for small values of the external dimensions, there is one unique isolated allowed region corresponding to the Ising model for all navigator functions that we study. In this case all searches are expected to lead to this same island.

For large values of the external dimensions on the other hand, it is known that generally numerical bootstrap bounds become very weak. Thus, the expectation is that, even under any of the sparseness assumptions studied here, there will also always exists some additional allowed "peninsula" for some very large values of the external dimensions. 

To minimize the navigator we use the modified BFGS algorithm proposed in \cite{Reehorst:2021ykw}. This is a quasi Newton's method where gradient information at different points is used to construct a convex quadratic model. At every iteration the navigator is first evaluated at the Newton's step to test whether this leads to sufficient decrease. If so the step is taken, the Hessian is updated with the new gradient information, and a new Newton's step is attempted and so on. If not, a line search is performed in the same direction as the Newton's step, until a point of sufficient decrease has been found. This direction is guaranteed to locally be a descent direction by construction. As input the algorithm takes an initial point and a bounding box describing the area that is to be searched. This bounding box also sets the initial scale of the Hessian. See algorithm 1 of \cite{Reehorst:2021ykw} for more details on this algorithm.%
\footnote{ 
    \cite{Reehorst:2021ykw}  also showed that in addition to the gradient the Hessian can also be computed cheaply. However, just like \cite{Reehorst:2021ykw}, this work still does not use the Hessian. Attempts to use it in the trust-radius based Newton method, \texttt{`trust-constr'} available in SciPy did not outperform the BFGS search. Possible causes are non-convexity of the navigator function and/or some non-$C^2$ differentiability in isolated places, see section 3.1 of \cite{Reehorst:2021ykw}. It could also be that the difference in performance lies in the strategies to deal with inaccuracies of the quadratic model. The SciPy BFGS algorithm uses a line search while its Newton's algorithms use trust-radius methods. Perhaps line search methods are better suited for the type of functions we encounter. Trust-radius methods update the step size based on the difference between the computed function value and the value predicted from its model. In the case of (Quasi) Newton's type algorithms a convex quadratic model is constructed from previous function and gradient calls or from a supplied Hessian. However, the navigator function is known to be non-convex at least when away from the minimum \cite{Reehorst:2021ykw}. (Minimizing  $\frac{\N(\bf{p})}{1-\N(\bf{p})}$ instead of the navigator directly, as suggested by \cite{Reehorst:2021ykw}, offers some improvement but even this function is generally non-convex). Therefore, the model will always be wrong in these regions and an inappropriate step size will be chosen. It is thus not surprising that these trust-radius methods have problems traversing these non-convex regions. A Newton's algorithm with a line search method might outperform a BFGS search. Unfortunately we did not find an easily available implementation of this. Therefore we leave this for future work.
} 

In order to choose the initial point and a suitable bounding box we tried multiple approaches. A non-rigorous EFM estimate was available \cite{Simmons-Duffin:2016wlq} for all parameters explored in this work. Starting from the median value of these estimates and using a bounding box that is a bit bigger than the error estimated by EFM leads to an allowed point in the estimated interval relatively quickly. We also tried a more agnostic approach where we start significantly away from the EFM estimates and search a wider bounding box. These tasks take longer to converge and, in all cases, we see that such a search either: ends in the same point as the other search; or, when the initial point is too far away from the EFM estimate, the search flows away to large values of the external dimensions where the navigator value becomes smaller and smaller and eventually becomes negative.

Traditional numerical bootstrap methods tried to map out and visualize the entire shape of the island. However, the number of function calls (or SDPB runs) required for these traditional methods scales exponentially with the dimensionality of the search space and would be infeasible for many of the searches presented in this work. Moreover, even the task of creating a meaningful visualizations becomes complicated in high-dimensional search spaces.

Instead, once an allowed point has been found, we use a constrained BFGS optimization algorithm (algorithm 2 from \cite{Reehorst:2021ykw}) to min or maximize some parameter $p_i$ under the constraint $\cN(\bf{p})\leq0$. That is we look for the locally minimal or maximal allowed values continuously connected to the initial allowed point. Repeating this for all parameters in both directions gives an $N$-dimensional hypercube in which the Ising CFT is constrained to live and thus gives rigorous bounds on the individual quantities in an easily quotable form. 

To solve the constraint optimization problem only a single initial allowed point is required. Still, we choose to complete the navigator minimization (algorithm 1) and to report the position of the minima. It was hypothesized in \cite{Reehorst:2021ykw} that the location of the minimum could be used as a predictor of the true location of the minimum. Thus, it is interesting to see whether the locations of the minima of different navigator functions are the same or different (for the subspace of parameters that the navigator functions in question have in common).

Note that we only solve a local optimization problem. This means that a smaller or greater allowed value could exist and remain unexplored if the island has a non-convex shape and we did not manage to probe some appendix of it. In that case our bounds would not be rigorous and a more difficult global optimization problem would have to be solved in order to find rigorous bounds. Similar problems occur when using traditional feasibility-based bootstrap methods. In that case one can also never exclude the possibility of the existence of a nearby missed tiny isolated allowed region nor the existence of a small appendix to the island that allows for smaller or larger allowed parameter values. 

Most (but not all) islands found in the numerical conformal bootstrap are found to be convex when a sufficiently constraining bootstrap setup is applied. In order to probe the possibility of a non-convex island shape and the existence of multiple locally extremal allowed values we ran all extremization tasks multiple times, starting from different allowed points within the island. In all cases, these different tasks lead to the same extremal values.\footnote{We also checked that the result of the minimization / maximization was indeed the minimal / maximal point found among all function calls in the various tasks.} Thus, we expect that the local extrema that we found are also globally the minimal and maximal allowed values (that are continuously connected to the initial allowed point).

We found additional sufficiently different allowed initial points for algorithm 2 in two ways. Firstly, we used points found in minimization runs (algorithm 1) that started from different points and thus approached the island from a different direction. Secondly we used allowed points found close to the minimal and maximal allowed value in the extremization in a different direction. We hope that starting from opposite sides of the island (in some parameter) maximizes our chance of finding all local minimal and maximal allowed values.

\section{Results}
\label{sec:results}

\subsection{The $\bZ_2$-even scalar sector}
\label{sec:results:even-scalar}
Minimizing the navigator $\N(\Delta_\epsilon,\Delta_\sigma,\theta,\Delta_{\epsilon'})$, starting from $\vec{p_0}=(0.510602$, $1.30533$, $0.888996$, $4.10374)$,  i.e. some random value in the vicinity of the Ising model, we find a minimum at $(0.5182848977$, $1.414807869$, $0.972072829$, $3.867504042)$ in 116 function calls. The first negative, and thus allowed, point is found on the 59-th function call. On the other hand starting from the median values obtained in \cite{Simmons-Duffin:2016wlq} the same minimum is found after only 66 function calls (in this case the initial point already has a negative navigator value). If the minimization algorithm starts too far away from the Ising model it instead flows to large values of the external dimensions.\footnote{This happens for example for the points $(0.525808$, $1.43445$, $0.531679$, $2.11166)$,  $(0.527961$, $1.43218$, $1.13252$, $2.01363)$ and $(0.512759$, $1.39774$, $1.11761$, $3.54061)$.} When we increase $\Lambda$ to 19 the minimum shifts to $(0.5181541969, 1.412697085, 0.969336983, 3.831157448)$, i.e. much closer to the EFM estimates that were previously obtained at $\Lambda=43$ (see Table~2 of \cite{Simmons-Duffin:2016wlq} or the last column of Table~\ref{table:bounds_7param}).

\begin{figure}[htpb]
\centering
\includegraphics[width = 0.45\textwidth]{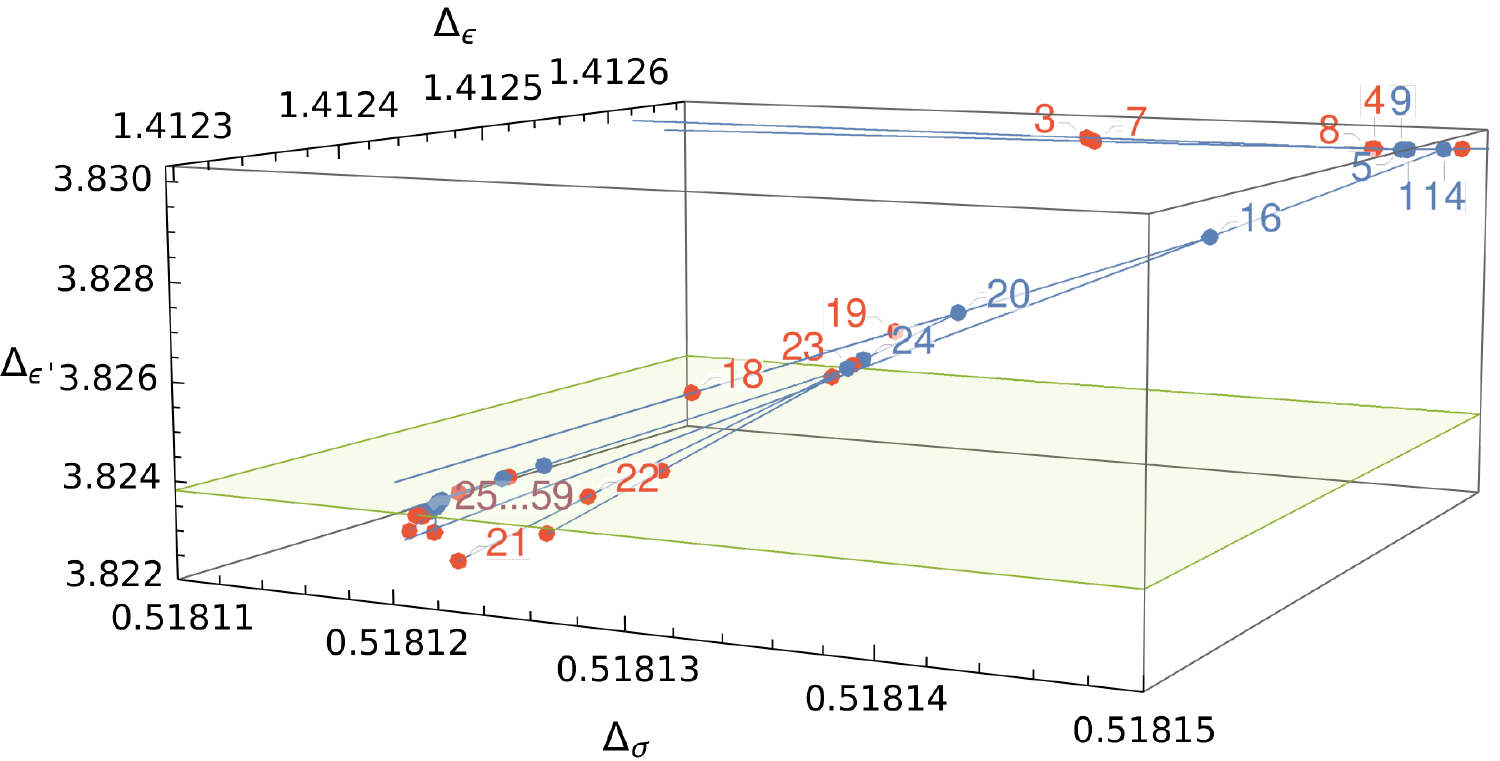}
\quad
\includegraphics[width = 0.45\textwidth]{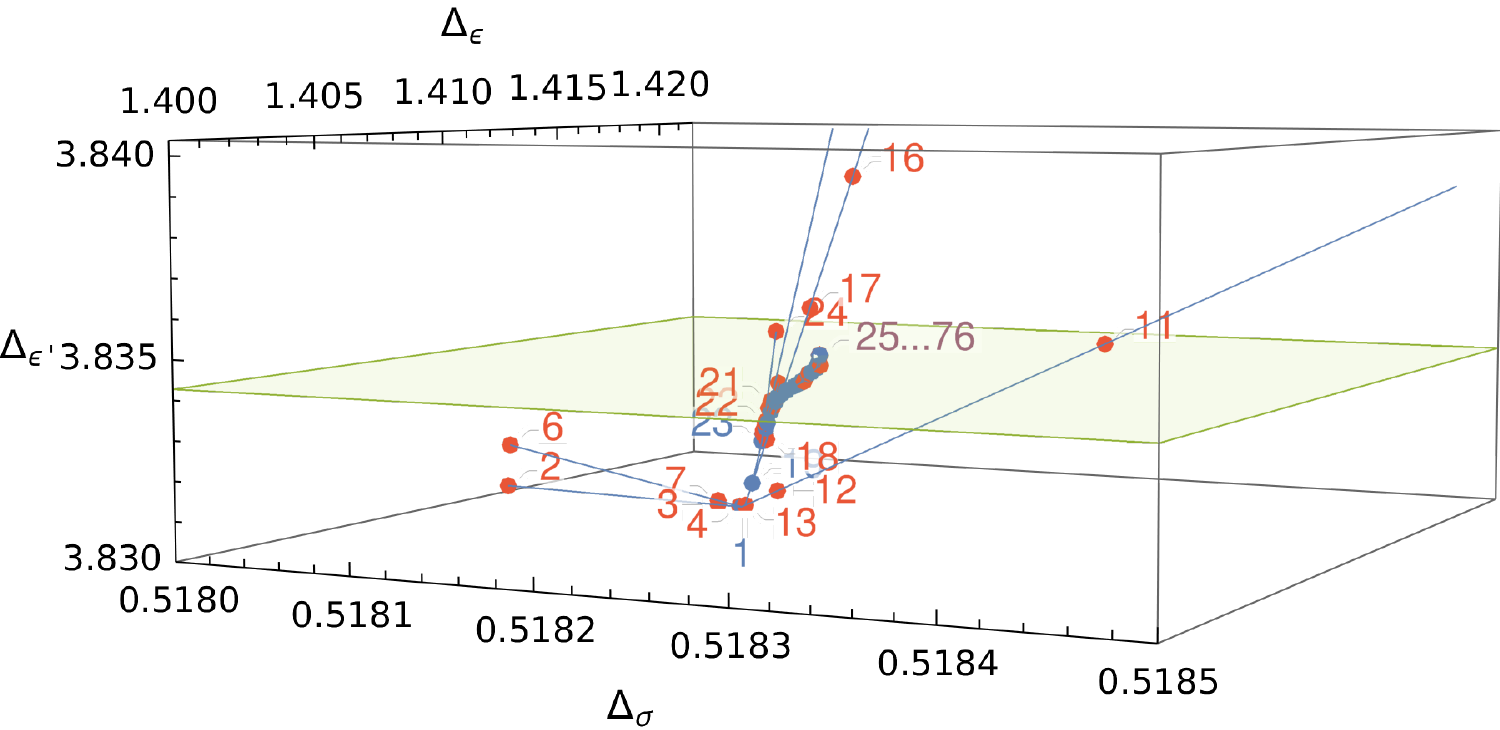}
\caption{\label{fig:example_extremization} Example runs of the parameter minimization/maximization algorithm in the 3d Ising island for the 
$\N(\Delta_\epsilon,\Delta_\sigma,\theta,\Delta_{\epsilon'})$ navigator (at $\Lambda=19$). The search is projected to the 3d subspace $(\Delta_\epsilon,\Delta_\sigma,\Delta_{\epsilon'})$. Blue points are allowed points (negative navigator value) and red points are disallowed (positive navigator value). The green sheet indicates the obtained minimal/maximal values.  \emph{Left:} Minimization of $\Delta_{\epsilon'}$ within the Ising island \emph{Right:}  Maximization of $\Delta_{\epsilon'}$ within the Ising island.}
\end{figure}

Given the allowed points found using algorithm~1, we used algorithm~2 to determine the minimal and maximal allowed values of the parameters. The resulting bounds at derivative orders $\Lambda=11$ and $19$ are shown in Table~\ref{table:bounds_4param}. An example visualization of two runs of algorithm~2 is shown in Figure~\ref{fig:example_extremization}.

The extremum was found on average after around 40 function calls for $\Lambda=11$ and about 60 function calls at $\Lambda=19$. In order to better compare the size of these rigorous bounds to the estimate obtained at very high derivative order of $\Lambda=43$ we also extremized in the $\Delta_{\epsilon'}$ directions at $\Lambda=31$ and found
\begin{equation}
    \Delta_{\epsilon'}=3.82951(\mathbf{61}) \qquad \textrm{(at $\Lambda=31$)} .
\end{equation}
The results obtained at $\Lambda=11$, $19$ and $31$ suggest that the rigorous bounds will converge to something very close to the EFM estimate $3.82968(23)$ \cite{Simmons-Duffin:2016wlq} where the "error bar" is given by the standard deviation of the values found in the EFM spectra obtained at $\Lambda=43$ for various points in the allowed Ising island.\footnote{Our setup for writing the SDP's has not yet been optimized for such a large scale bootstrap problem so we leave the comparison of results obtained at $\Lambda=43$ in both methods to a future work.}
\begin{table*}[h!]
	\centering
	\begin{tabular}{@{}l l l l l @{}}
		\toprule
		& $\Lambda=11$  bounds & $\Lambda=11$  minimum & $\Lambda=19$ bounds & $\Lambda=19$ minimum \\
		\midrule
		 $\Delta_\sigma$& $0.51866(\mathbf{93})$ & $0.51828489773601$ & $0.518157(\mathbf{35})$ & $0.51815419690124$ \\
		$\Delta_\epsilon$  & $1.4156(\mathbf{79})$ & $1.41480786868088$ & $1.41265(\mathbf{36})$ & $1.41269708524225$\\
		$\theta$ & $0.9701(\mathbf{43})$ & $0.97207282901339$ & $0.96925(\mathbf{23})$ & $0.96933698299715$\\
		$\Delta_{\epsilon'}$ & $3.841(\mathbf{76})$ & $3.86750404211864$ &$3.8290(\mathbf{52})$ & $3.83115744805532$\\
		\bottomrule 
	\end{tabular}
	\caption{Allowed interval found at $\Lambda=11$ and $19$ using the $\N(\Delta_\epsilon,\Delta_\sigma,\theta,\Delta_{\epsilon'})$ navigator where we assume $\Delta_{\sigma'}>3$ and $\Delta_{\epsilon''}>6$. The error presented in bold is rigorous. The second and fourth column give the location of the minimum at $\Lambda=11$ and $19$ respectively.
	}
	\label{table:bounds_4param}
\end{table*}

We can examine the same question with a navigator function $\N(\Delta_\epsilon$, $\Delta_\sigma$,$\lambda_{\sigma \sigma \epsilon}$, $\lambda_{\epsilon \epsilon \epsilon}$, $\Delta_{\epsilon'}$, $\lambda_{\sigma \sigma \epsilon'}$, $\lambda_{\epsilon \epsilon \epsilon'})$ where we also scan over all OPE coefficients of the two isolated lowest dimensional $\bZ_2$-even scalars. Starting from $\mathbf{p}_0=(0.51819$, $1.41301$, $1.05185$, $1.53244$, $3.83427$, $0.053012$, $1.536)$ the first negative point is found after $89$ function calls and the minimum is reached after $235$ function calls. The bounds found at $\Lambda=19$ are shown in table~\ref{table:bounds_7param}. 

When non-degeneracy at the dimension $\Delta_{\epsilon}$ is imposed by contracting with an angle $\theta$ this significantly improves bounds \cite{SlavaUnpublished,Kos:2016ysd}. Instead here we see that there is no noticeable change in the bounds on $\Delta_\epsilon$, $\Delta_\sigma$ or $\Delta_{\epsilon'}$ when we scan over the OPE coefficients of $\epsilon'$.\footnote{Likely we would have to consider the larger bootstrap setup involving the correlators of $\epsilon'$ to see this assumption have a stronger effect.}
\begin{table*}[h!]
	\centering
	\begin{tabular}{@{}l l l l@{}}
		\toprule
		&  $\Lambda=19$ bounds & $\Lambda=19$ minimum & EFM estimate \cite{Simmons-Duffin:2016wlq}\\
		\midrule
		$\Delta_\sigma$ &$0.518157(\mathbf{35})$ & $0.5181541969$ & $0.5181489(\mathbf{10})$\\
	    $\Delta_\epsilon$ & $1.41265(\mathbf{36})$ & $1.412697085$ & $1.412625(\mathbf{10})$\\
		$\lambda_{\sigma\sigma\epsilon}$ & $1.05185(\mathbf{12})$ & $1.051827659$& $1.0518537(41)$\\
		$\lambda_{\epsilon\epsilon\epsilon}$ & $1.53240(\mathbf{58})$ & $1.532647927$ & $1.532435(19)$\\
		$\Delta_{\epsilon'}$& $3.8290(\mathbf{52})$ & $3.831157444$ & $3.82968(23)$\\
		$\lambda_{\sigma\sigma\epsilon'}$ &$0.05304(\mathbf{16})$ & $0.05298356893$ & $0.053012(55)$\\
		$\lambda_{\epsilon\epsilon\epsilon'}$ & $1.5362(\mathbf{12})$ & $1.536468123$ & $1.5360(16)$\\
		\bottomrule 
	\end{tabular}
	\caption{Allowed interval found at $\Lambda=11$ and $19$ using the $\N(\Delta_\epsilon,\Delta_\sigma,\lambda_{\sigma\sigma\epsilon}$,$\lambda_{\epsilon\epsilon\epsilon}$,$\Delta_{\epsilon'}$,$\lambda_{\sigma\sigma\epsilon'}$,$\lambda_{\epsilon\epsilon\epsilon'})$ navigator where we assume $\Delta_{\sigma'}>3$ and $\Delta_{\epsilon''}>6$ and that there is no degeneracy at $\Delta_{\epsilon'}$. The error presented in bold is rigorous. The EFM estimates obtained in \cite{Simmons-Duffin:2016wlq} are included for reference.
	}
	\label{table:bounds_7param}
\end{table*}

\begin{figure}[htpb]
\centering
\includegraphics[scale=0.8]{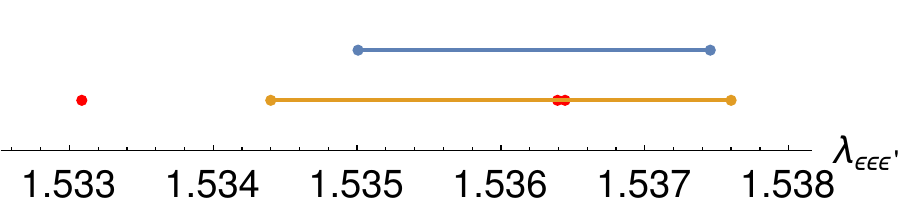}
\caption{\label{fig:allowed_interval_OPE1} In blue: Rigorous bounds on $\lambda_{\epsilon\epsilon\epsilon'}$ found at $\Lambda=19$ using algorithm 2. In orange: The interval corresponding to one standard deviations of the values found for $\lambda_{\epsilon\epsilon\epsilon'}$ using the extremal functional as found in \cite{Simmons-Duffin:2016wlq} at $\Lambda=43$. The red dots indicate the values we found using \texttt{spectrum} at $\Lambda=43$ under the assumptions the assumptions $\Delta_{\e'}>3$ and $\Delta_{\s'}>3$ for the points $(0.51814898, 1.4126250, 0.9692610)$, $(0.51814898, 1.412625, 0.969261 )$ and $(0.51814931, 1.4126302, 0.9692658)$ (in order of increasing $\lambda_{\e\e\e'}$). Note that the first point gave a spectrum with a $\lambda_{\epsilon\epsilon\epsilon'}$ value excluded by our rigorous bounds (despite the bound having been derived at a lower value of $\Lambda$). This can be explained by the \emph{sharing effect} \cite{Liu:2020tpf}; see the main text.
}
\end{figure}

However, surprisingly the rigorous bounds on $\lambda_{\epsilon \epsilon \epsilon'}$ are slightly stronger than the error bars given by the standard deviation of values found by EFM method (despite using a much lower derivative order here). This is illustrated in figure~\ref{fig:allowed_interval_OPE1}. 

It was noted in \cite{Liu:2020tpf} that the extremal functional method seems less accurate for OPE coefficients possibly due to the so called \emph{sharing effect} where false contributions are coming from fake operators at or close to the imposed gaps. By implicitly scanning over $\Delta_{\epsilon'}$ we should be avoiding these contributions which could explain why we can exclude part of the error bar of the EFM result.

To investigate this we repeated the EFM computation at $\Lambda=43$ in \cite{Simmons-Duffin:2016wlq} for three allowed points (one in the middle and two near two edges), using the assumptions $\Delta_{\e'}>3$ and $\Delta_{\s'}>3$. Interestingly, at the point $(0.51814931, 1.4126302, 0.9692658)$ we found an example of an extremal function for which \texttt{spectrum} \cite{DSD-spectrum-extraction} outputs a spectrum with a value for $\lambda_{\epsilon \epsilon \epsilon'}$ below those allowed by our rigorous bounds.\footnote{The current version of \texttt{spectrum} has a bug where the OPE contribution for a constant vector is not outputted. To ensure that the outputted spectrum was otherwise correct we also compared it against the older code \texttt{spectrum.py}, which can be found at \cite{DSD-spectrum-extraction}. Note that \texttt{spectrum.py} is incompatible with the in and output formats used by current versions of SDPB.}
However, in this spectrum we also see a contribution from an operator at $\Delta_\epsilon=3$ with sizeable OPE coefficients, which we are forbidding in our navigator search with our assumption $\Delta_{\epsilon''}>6$.\footnote{Naively adding the OPE coefficients $\lambda_{\epsilon \epsilon \epsilon'}$ and $\lambda_{\epsilon \epsilon \cO_{\Delta=3}}$ together for this spectrum instead gives an OPE value above our rigorous bounds. We did not find any spectrum with a value for $\lambda_{\epsilon \epsilon \epsilon'}$ above our rigorous bounds. A value that is too high would be harder to explain by the presence of spurious operators. Note that \cite{Simmons-Duffin:2016wlq} only provides the median value and the (symmetric) standard deviation. For a skewed distribution it is entirely possible that no value at the upper end of the one standard deviation interval was actually obtained.}

We thus expect the mismatch to be due to an error in the value of $\lambda_{\e\e\e'}$ obtained by extremal functional method and we expect our rigorous bounds to be correct.

Finally we tested the dependence on the exact gap that we imposed. We find that the exact position of the gap does not have a strong effect on the bounds that were found. In Table~\ref{table:bounds_4param_gap_test} we compare the bounds we obtained using the gap assumptions $\Delta_{\e''}>5.5, 6, 6.5$. The upper bound remains exactly stable while the lower bound shows only some very minor dependence on the imposed gap. Note also that when we considering the navigator function $\N(\Delta_{\sigma},\Delta_{\epsilon},\theta)$ with a gap at 6 without allowing for the presence of an operator with dimension $\Delta_{\e'}$ below it we do not find any allowed region for small values of $\Delta_{\sigma}$ and ${\Delta_\epsilon}$. Instead the navigator-search tends towards large $\Delta_{\sigma}$ and ${\Delta_\epsilon}$ values where it eventually becomes negative. As expected there does not seem to exist a theory with only a single $\bZ_2$ even scalar below a gap of 6 for small values of $\Delta_{\sigma}$ and $\Delta_{\epsilon}$.\footnote{The same holds for the other sectors we studied. I.e. if we assume $\Delta{T'}>6$ or $\Delta_{\s'}>6$ no allowed Ising island can be found and the navigator search instead flows towards large values of the external operator dimensions.}

\begin{table*}[h!]
	\centering
	\begin{tabular}{@{}l l l l@{}}
		\toprule
		& $\Delta_{\e''}>5.5$ & $\Delta_{\e''}>6$  & $\Delta_{\e''}>6.5$ \\
		\midrule
		$\Delta_\sigma$ & $0.518156(\mathbf{36})$ & $0.518157(\mathbf{35})$ & $0.518159(\mathbf{33})$ \\
		$\Delta_\epsilon$  & $1.41264(\mathbf{37})$ & $1.41265(\mathbf{36})$ & $1.41268(\mathbf{33})$\\
		$\theta$ & $0.96924(\mathbf{24})$ & $0.96925(\mathbf{23})$ & $0.96927(\mathbf{21})$\\
		$\Delta_{\epsilon'}$ & $3.8286(\mathbf{57})$ & $3.8290(\mathbf{52})$ &$3.8297(\mathbf{45})$\\
		\bottomrule 
	\end{tabular}
	\caption{The three columns show the allowed interval found at $\Lambda=19$ under the assumptions $\Delta_{\e''}>5.5, 6$ and $  6.5$ respectively. The error presented in bold is rigorous. The upper bound is completely stable under a change of the position of the gap while the lower bound shows some very minor change.
	}
	\label{table:bounds_4param_gap_test}
\end{table*}

\subsection{The $\bZ_2$-odd scalar sector}
\label{sec:results:odd-scalar}
When we instead assume $\Delta_{\epsilon'}>3$ and $\Delta_{\sigma''}>6$, i.e. the existence of exactly two $\bZ_2$-odd operators below 6, we again find a single isolated allowed region near the expected position of the Ising model. Minimizing the $\N(\Delta_\epsilon,\Delta_\sigma,\theta,\Delta_{\sigma'})$ navigator, starting for example from $\mathbf{p}_0=(0.518148, 1.41262, 0.969, 4.5)$ the minimum at $(0.5181541678$, $1.412696437$, $0.9693359032$, $5.299966347)$ is reached after $204$ iterations. Here we started quite far away in the $\Delta_{\sigma}$ coordinates and still converge to the Ising model CFT successfully. Also in this case we see that starting from the EFM estimate, which is relatively close to the real value of the minimum, the number of function calls is roughly cut in half, in this case to $111$. Extremizing in the $\Delta_{\sigma'}$ direction we find the following bounds
\begin{equation}
    \begin{split}
        \Delta_{\sigma'}=5.262(\mathbf{89}) \qquad \textrm{(at $\Lambda=19$)} .
    \end{split}
\end{equation}
In table~\ref{table:bounds_4param_sigma} below we also show the other bounds (as well as the location of the minimum). From these bounds it can be seen that the sparseness conditions $\Delta_{\epsilon''}>3$ and $\Delta_{\sigma''}>6$ are a bit less constraining than the "reverse" assumptions $\Delta_{\epsilon''}>6$ and $\Delta_{\sigma''}>3$ studied in the previous section. It is also clear that the two minima don't coincide in the common $(\Delta_\epsilon,\Delta_\sigma,\theta)$ parameter space. This somewhat goes against the hypothesis that there is a special meaning or great predictive value to the exact position of the minimum of a certain navigator function at a finite value of $\Lambda$. 
\begin{table*}[h!]
	\centering
	\begin{tabular}{@{}l l l@{}}
		\toprule
		&  $\Lambda=19$ bounds & $\Lambda=19$ minimum \\
		\midrule
		$\Delta_\sigma$ &  $0.518153(\mathbf{40})$ & $0.5181640959254$ \\
		$\Delta_\epsilon$  &  $1.41260(\mathbf{41})$ & $1.4128014381733$\\
		$\theta$ &  $0.96920(\mathbf{28})$ & $0.9693835270248$\\
		$\Delta_{\sigma'}$ &$5.262(\mathbf{89})$ & $5.1725929510462$\\
		\bottomrule 
	\end{tabular}
	\caption{Allowed interval and minimum found at $\Lambda=19$ using the $\N(\Delta_\epsilon,\Delta_\sigma,\theta,\Delta_{\sigma'})$ navigator where we assume $\Delta_{\sigma''}>6$ and $\Delta_{\epsilon'}>3$. The error presented in bold is rigorous. 
	}
	\label{table:bounds_4param_sigma}
\end{table*}

\sloppy If we also choose to scan over the OPE coefficient $\lambda_{\sigma\epsilon \sigma'}$ we find a minimum of $\N(\Delta_\epsilon,\Delta_\sigma,\theta,\Delta_{\sigma'},\lambda_{\sigma\epsilon \sigma'})$ at $(0.5181541678$, $1.412696437$, $0.9693359032$, 
$5.299966347$, $0.0572226234)$. Extremizing in the $\lambda_{\sigma\epsilon\sigma'}$ directions we find  
\begin{equation}
    \begin{split}
        \lambda_{\sigma\epsilon\sigma'}=0.0565(\mathbf{15}) \qquad \textrm{(at $\Lambda=19$)} .
    \end{split}
\end{equation}
Again this closely matches the value $0.057235(20)$ obtained by EFM estimate and we expect that the rigorous bounds will converge to a similar interval at $\Lambda=43$.

\subsection{The $\bZ_2$-even spin-2 sector}
\label{sec:results:even-spin2}
In the spin-2 $\bZ_2$-even sector one quantity of physical interest is the central charge $c_T$ related to the stress tensor OPE coefficients. In order to find an isolated allowed interval in $c_T$ we need to assume some gap on the allowed dimension of the next $\bZ_2$-even spin-2 operator. For example, using $\Delta_{T'}>4$ we find, using $\N(\Delta_\epsilon,\Delta_\sigma,\theta,c_T)$, that the central charges must take a value inside the interval $\frac{c_T}{c_T^{\textrm{free}}}=0.946544(\mathbf{43})$. Instead we can also assume that there are only two spin-2 operators with a dimension below 6. This assumption on the sparseness of the spin-2 spectrum leads to a slightly stronger bound on $c_T$ and also allows us to bound $\Delta_{T'}$ at the same time. Using $\N(\Delta_\epsilon,\Delta_\sigma,\theta,c_T,\Delta_{T'})$ we find: 
\begin{equation}
    \begin{split}
        \frac{c_T}{c_T^{\textrm{free}}} &= 0.946543(\mathbf{42}) \qquad \textrm{(at $\Lambda=19$)} \\
         \Delta_{T'}&=5.499(\mathbf{17}) \qquad \hspace{17pt} \textrm{(at $\Lambda=19$)}.
    \end{split}
\end{equation}
Again, this bound matches the expectations from  the EFM spectrum as well the estimate that was obtained in \cite{El-Showk:2014dwa} by assuming that the Ising model lives close to the kink in the lower bounds on $c_T$ as a function of $\Delta_\sigma$.

Again we can also include all the OPE coefficients including $\Delta_{T'}$ in the navigator search. The minimum of the resulting 8 parameter navigator function $\N(\Delta_\epsilon,\Delta_\sigma,\lambda_{\s\s\e},\lambda_{\e\e\e},c_T,\Delta_{T'}, \lambda_{\s\s T'},\lambda_{\e\e T'})$ is found to be at $(0.5181458953$, $1.412593006$, $1.051864373$, $1.532394006$, $1.58471279$, $5.510997316$, $0.02114139175$, $1.382481992)$ and we find the bounds
\begin{equation}
    \begin{split}
        \lambda_{\s\s T'} &= 0.02107(\mathbf{20}) \qquad \textrm{(at $\Lambda=19$)} \\
         \lambda_{\e\e T'}&=1.355(\mathbf{30}) \qquad \hspace{17pt} \textrm{(at $\Lambda=19$)}.
    \end{split}
\end{equation}
Note the factor of 2 difference with the EFM estimate in \cite{Simmons-Duffin:2016wlq} due to the use of a different conformal block normalization.

Looking at all the minima we found for the various navigator function we see that they do not necessarily coincide within the shared parameter spaces. Moreover, we do not see a navigator minimum that is clearly a better predictor for the location of the $\Lambda=43$ Ising island than the others. The closest minimum in the universal $(\Delta_{\s},\Delta_{\e})$ subspace occurs at $\mathbf{p}=(0.51814715, 1.412608214, 1.051859195, 1.532426026, 1.584714379)$ for the $\N(\Delta_\epsilon,\Delta_\sigma,\theta,c_T,\Delta_{T'})$ navigator. Surprisingly adding the scan over the OPE coefficients in $\N(\Delta_\epsilon,\Delta_\sigma,\lambda_{\s\s\e},\lambda_{\e\e\e},c_T,\Delta_{T'}, \lambda_{\s\s T'},\lambda_{\e\e T'})$ drives the minimum further away from the $\Lambda=43$ island rather than closer.

\subsection{Combined sparseness assumptions on all sectors}
Here we bound the same quantities studied above once more but now by simultaneously demanding all the sparseness conditions considered above. One could hope that imposing sparseness in all these sectors at once could be highly constraining leading to a major reduction of the size of the island. This problem results in a navigator function $\N(\Delta_\epsilon,\Delta_\sigma,\lambda_{\sigma \sigma \epsilon},\lambda_{\epsilon \epsilon \epsilon},\Delta_{\epsilon'},\lambda_{\sigma \sigma \epsilon'},\lambda_{\epsilon \epsilon \epsilon'},\Delta_{\sigma'},\lambda_{\sigma\epsilon \sigma'},c_T, \Delta_{T'},\lambda_{\sigma\sigma T'},\lambda_{\epsilon\epsilon T'})$ that depends on 13 parameter.  

Even given the EFM results the task of finding an allowed point in this space is very non-trivial. There are a few methods we could use to find a decent initial guess to start our navigator search from. Firstly, we can work iteratively. We can take the allowed or even minimal points that we found when scanning over a smaller subset of data and start our navigator search from the resulting point. Secondly, we can take the \emph{median EFM spectrum}, i.e. the spectrum constructed by taking the median values obtained by EFM method presented in Table 2 of \cite{Simmons-Duffin:2016wlq}. 

Neither of these initial guesses will immediately provide an allowed point. The combined sparseness conditions are more constraining than any of the individual ones. A solution found in a single extremal spectrum at $\Lambda=43$ should give an allowed point $\bf{p}$ at any lower value of $\Lambda$. However, a point $\bf{p}_{\textrm{med}}$ constructed from the median value of multiple spectra does not have to give a solution. And indeed we find it does not. This can be explained by the island having a thin elongated shape as opposed to being a hyper-cube or hyper-ball. Locally the island can be thin in certain directions so even a small error can put a point outside the allowed space. We will revisit this issue at the end of this section.

Finally, we could try to find an initial allowed point for our navigator search by starting from the extremal spectrum extracted from a single point in the previously found $\Lambda=43$ island. If this spectrum obeys the spectrum assumptions we make here it should instantly provide an allowed point. However, when we repeated the EFM computation for a few points at $\Lambda=43$ we found they contained either "spurious zeros" which were excluded from Table 2 of \cite{Simmons-Duffin:2016wlq} by hand and which do not obey our sparseness assumptions or they contained inaccurate OPE values due to the \emph{sharing effect} (see Figure~\ref{fig:allowed_interval_OPE1} and the related comments in Section~\ref{sec:results:even-scalar}). No spectrum gave an allowed point once we removed the spurious contributions inconsistent with our sparseness assumptions even at the much lower derivative order $\Lambda=19$.

Here we present the case where we start from the \emph{median EFM spectrum}. This  leads to an allowed point faster than starting from an initial point constructed from the minimal values found by imposing sparseness on the various sectors one by one. Moreover, it involves less arbitrary choice for how to construct the initial point. In this case a negative, and thus allowed, point was found after 201 function calls. The convergence to the exact minimum is much slower for the 13 parameter navigator function than for any of the other navigator functions we studied. Non-convexity seems to play a major role in this. We will comment more on this in Section~\ref{sec:convergence}. As a result the minimization has not fully converged even after over 400 function calls.\footnote{The norm of the step size has converged to below $\cO(10^{-7})$ but the norm of the gradient is still of order $\cO(10^{-3})$.}

\begin{figure}[htpb]
\centering
\includegraphics[width=0.9\linewidth]{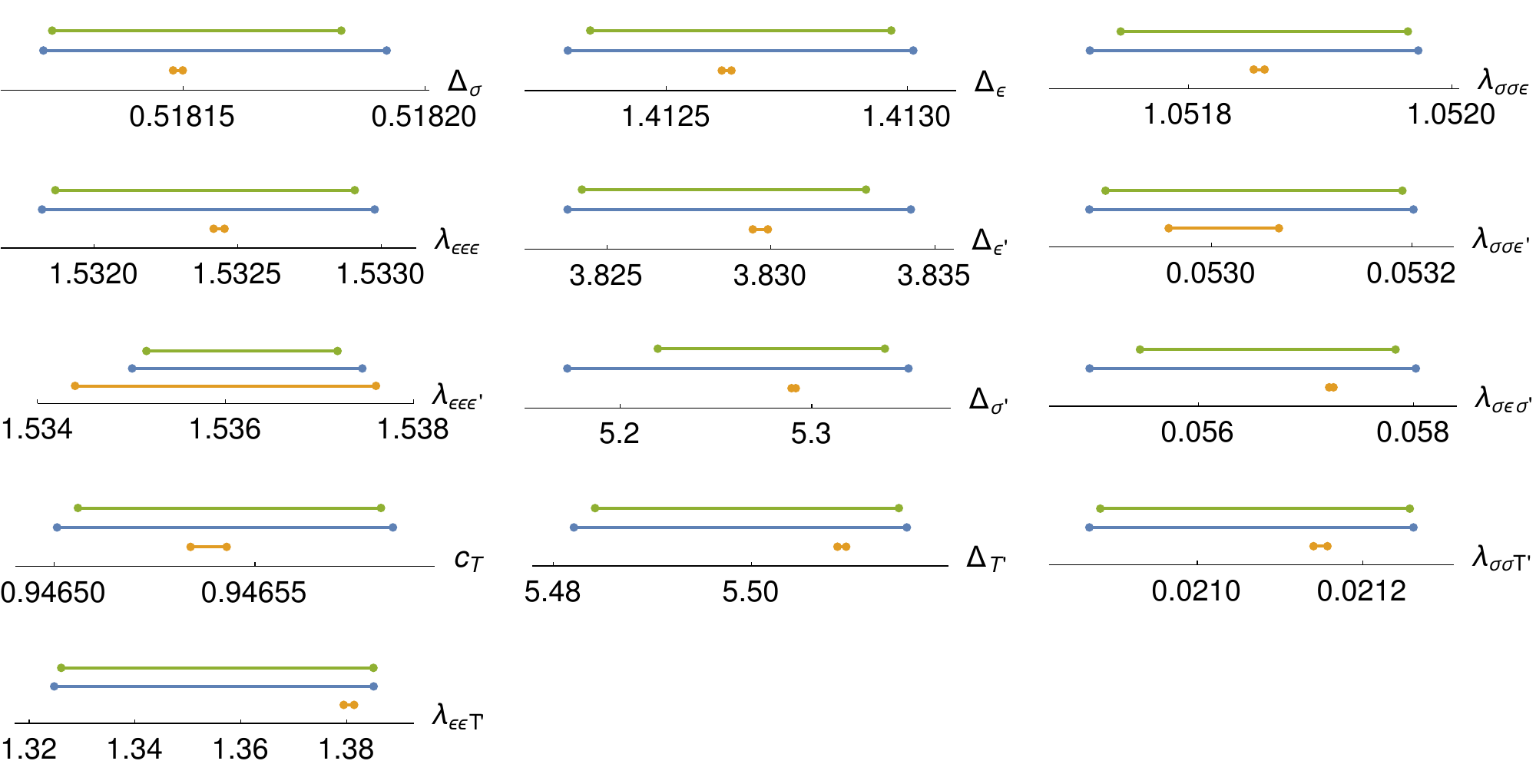}
\caption{\label{fig:13parambounds} Bounds on the OPE data for operators with a dimension below 6 for the sectors studied in this paper. In blue: Rigorous bounds found at $\Lambda=19$ by by demanding sparseness assumption of the form $\Delta_{\cO''}>6$ in one sector at a time. In green: The allowed interval found using the 13 parameter navigator function at $\Lambda=19$ using a sparseness assumption of the form $\Delta_{\cO''}>6$ for all sectors at once. The combined sparseness assumptions (only) slightly reduce the size of the allowed intervals.  The one-standard-deviation interval found using the EFM method at $\Lambda=43$ is included in orange for reference.
}
\end{figure}

Likewise the convergence of the parameter minimization/maximization is also much slower. The bounds we give in this section have not fully converged and so don't have the same rigor as the other bounds we presented in earlier sections.\footnote{ It is not certain that we found a point close to the local extremum because the gradient components orthogonal to the extremization direction are still of order O(1). However, towards the end of the constrained BFGS optimizations the step sizes have already converged to steps smaller than 0.01\% of the size of the bounds. Thus, we don't expect a noticeable difference between these bounds and the fully converged ones.}  We show these bounds pictorially in Figure~\ref{fig:13parambounds}. The main conclusion of this sections does not rely on the full convergence of these extremizations since it only depends on the existence of the allowed points we did already find. Based on those we can conclude that imposing sparseness assumption on all sectors at once does not greatly reduce the size of the bounds compared to bounds obtained applying the sparseness assumption to one single sector at a time. Of course, in the case of the Ising island the assumption that there is only one $\bZ_2$ even and odd scalar already isolates the Ising island sufficiently so that the extremal spectra for points within the island mostly obey the sparseness conditions we studied. These points cannot be excluded by the sparseness assumptions we impose (unless they contain "spurious zeros" as it turns out is often the case). Perhaps imposing sparseness in multiple sectors could still be effective to aid in isolating a physical theory in cases where lesser assumptions do not sufficiently isolate an island.\footnote{See for example the bootstrap of CFTs with $O(N)\cross O(M)$ symmetry \cite{Henriksson:2020fqi} or the bootstrap of the $ARP^3$ model \cite{Reehorst:2020phk}.}

\begin{figure}[htpb]
\centering
\includegraphics[width=0.9\linewidth]{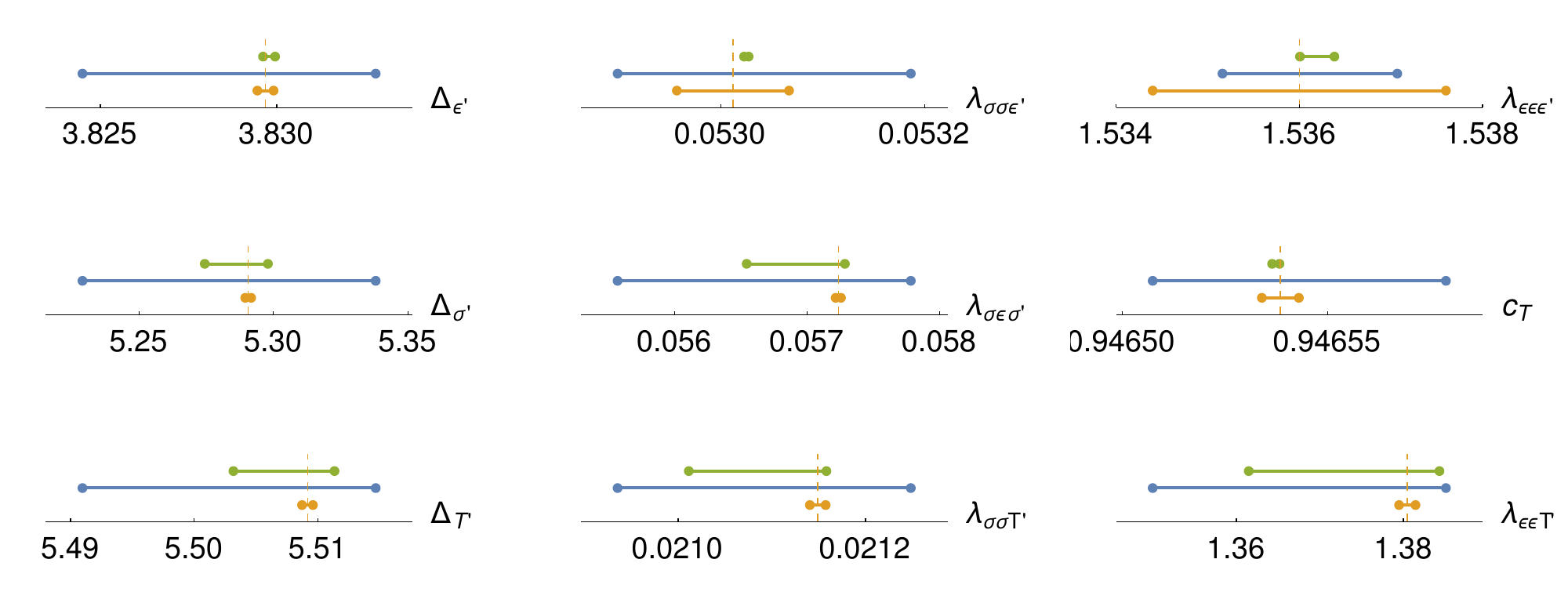}
\caption{\label{fig:locally_allowed_interval_fixed4} In green: the allowed interval after fixing the first 4 parameters to be $(\Delta _{\sigma },\Delta _{\epsilon },\lambda _{\sigma \sigma \epsilon },\lambda _{\epsilon \epsilon \epsilon })=(0.5181496671, 1.412636636, 1.05184946, 1.53247348)$. All allowed intervals shrink significantly when we do not allow movement in the first 4 parameters. The allowed intervals for $\lambda _{\sigma \sigma \epsilon '}$ and $c_T$ are especially small locally compared to the size of the full island. In blue: Rigorous bounds found at $\Lambda=19$ obtained by demanding sparseness assumption of the form $\Delta_{\cO''}>6$ in one sector at a time. In orange: The one-standard-deviation interval found using the EFM method at $\Lambda=43$ (included for reference). The median value is indicated by an orange dashed line. Note that this point lies outside the green interval for $\lambda _{\sigma \sigma \epsilon '},\lambda _{\epsilon \epsilon \epsilon '}$ and $c_T$.
}
\end{figure}

To also give an idea of the local shape of the island let's fix the first 4 parameters (i.e. the previously known ones) and find the minimal and maximal allowed values in the other parameters when we do not allow movement in the first 4.  We pick the first 4 parameters to be $(\Delta _{\sigma },\Delta _{\epsilon },\lambda _{\sigma \sigma \epsilon },\lambda _{\epsilon \epsilon \epsilon })=(0.5181496671, 1.412636636, 1.05184946, 1.53247348)$. These values are those of an allowed point found closes to the \emph{median EFM spectrum}.\footnote{Remember that the \emph{median EFM spectrum} itself does not give an allowed point. Instead this allowed point was found using the navigator minimization algorithm.} The resulting bounds are shown pictorially in Figure~\ref{fig:locally_allowed_interval_fixed4}. The figure shows that for certain variables the allowed interval shrinks significantly if we do not allow movement in the first 4 parameters. It also shows why the \emph{median EFM spectrum} does not give an allowed point itself since for the variables $\lambda _{\sigma \sigma \epsilon '}$,$\lambda _{\epsilon \epsilon \epsilon '}$ and $c_T$ the median EFM value falls outside the allowed interval (at these fixed values of the first 4 parameters). 

\subsection{Convergence and scaling of navigator searches}
\label{sec:convergence}
\begin{figure}[htpb]
\centering
\includegraphics[width = 0.45\textwidth]{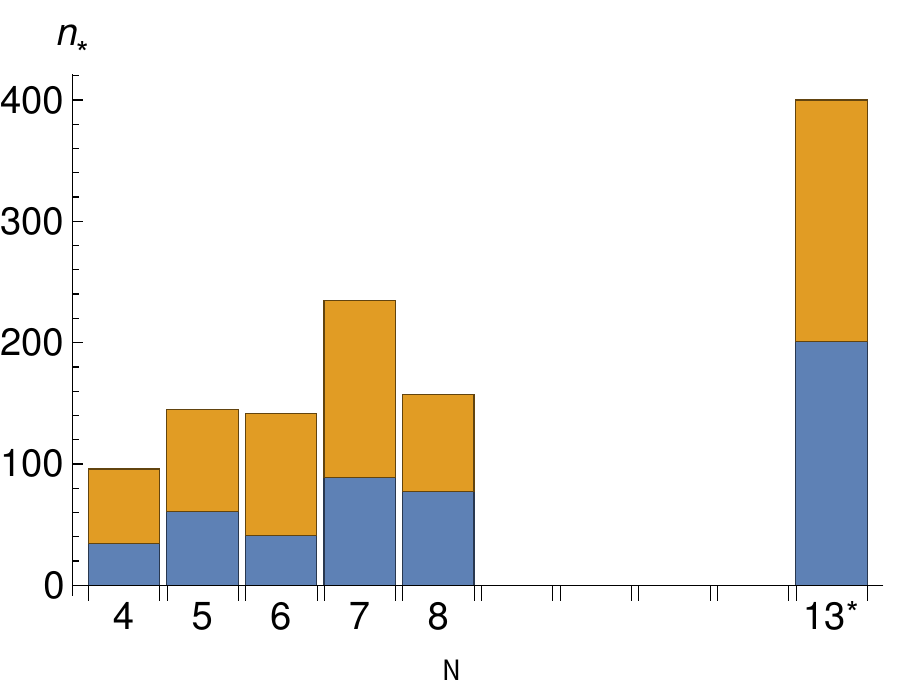}
\quad
\includegraphics[width = 0.45\textwidth]{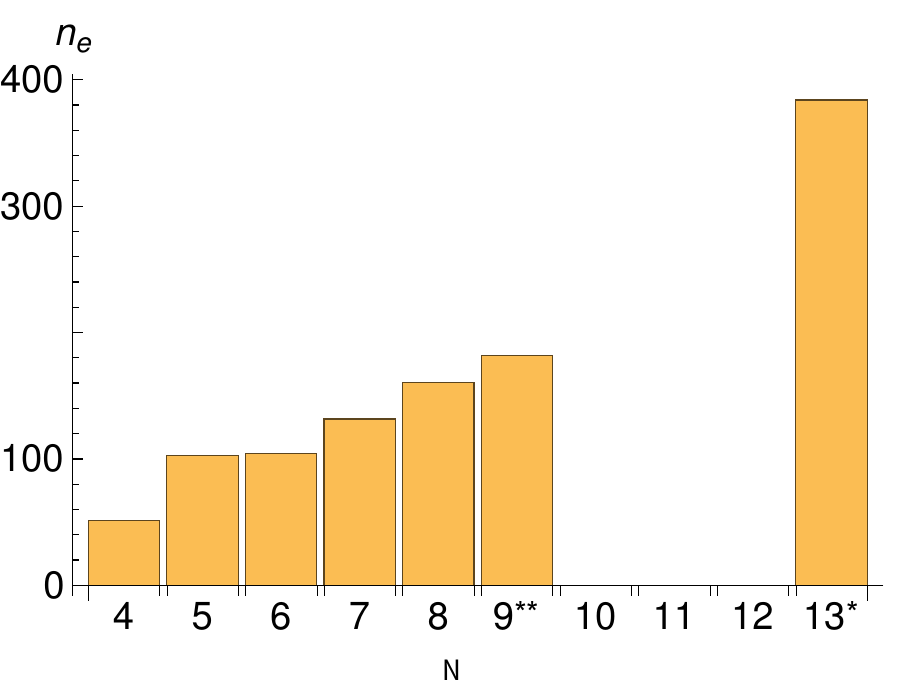}
\caption{\label{fig:ConstrBFGS_path} On the left: A bar chart showing the mean required number of function calls to find (in blue) an allowed point and (in orange) the minimum of the navigator function for parameter spaces of various dimensions $N$. On the right: The mean number of function calls required to find the minimal or maximal allowed value for one of the navigator function parameters in a search space of dimension $N$. \textasteriskcentered: Note that in the case of the 13 dimensional optimization problems the gradient threshold termination condition was never fully reached. However, the step sizes have converged to steps smaller than $\cO(10^{-7})$. \textasteriskcentered\textasteriskcentered: The 9 parameter optimization is the optimization of the 13-parameter navigator restricted to a 9 dimensional subspace by fixing 4 parameters to fixed values.}  
\end{figure}

In this section we look at the required number of function calls for navigator searches of spaces of different dimensionality. Of course, the required number of function calls depends strongly on the exact shape of the navigator function for a given problem and on the initial information that is available. Starting from a good guess and inputting an appropriate bounding box for the search space can significantly speed up the search. For some tasks we tried starting far away with wide assumed bounds to see that we still flow to the same point while for others we gave the best possible estimates and bounds. Nevertheless we present in Figure~\ref{fig:ConstrBFGS_path} (on the left) the naive mean number of function calls required to find an allowed point and the number of function calls required to find the minimum. This can still give some order of magnitude estimate of the required number of function calls for a generic navigator minimization.

We see that the required number of points scales more or less linearly with the dimensionality of the search space and definitely scales much better than the exponential scaling exhibited by traditional methods.

In 
the same figure, on the right, we also show the average number of function calls required to find the minimal and maximal allowed parameter continuously connected to a given initial allowed point.\footnote{ 
We end the extremization once the following termination criteria have been met:
\begin{equation}
\label{maxx1problemequivalent}
    \abs{\cN(x)} \leq g_{\text{tol}}=10^{-20} \qquad\text{and}\qquad
    \norm{\left(I- \frac{n n^T}{n^Tn}\right) \nabla \, \cN(x)}  \leq g_{\text{tol,g}}=10^{-10}\,,
\end{equation}
Note that when the navigator function as well as the gradient components orthogonal to the direction that is being extremized both tend to zero it means that locally a larger/smaller parameter value cannot be obtained without increasing the navigator function away from zero. This indicates the presence of (at least) a local extremum of the allowed island.
}
When extremizing a parameter within the island the choice of the initial point should matter less for the mean number of required steps since the same initial points were used for both parameter minimization and maximization. If the initial point is close to one boundary it should be far from the opposite boundary. Again we see a more or less linear dependence on the dimensionality of the search space.

\begin{figure}[htpb] 
\begin{subfigure}{0.48\textwidth}
\includegraphics[width=\linewidth]{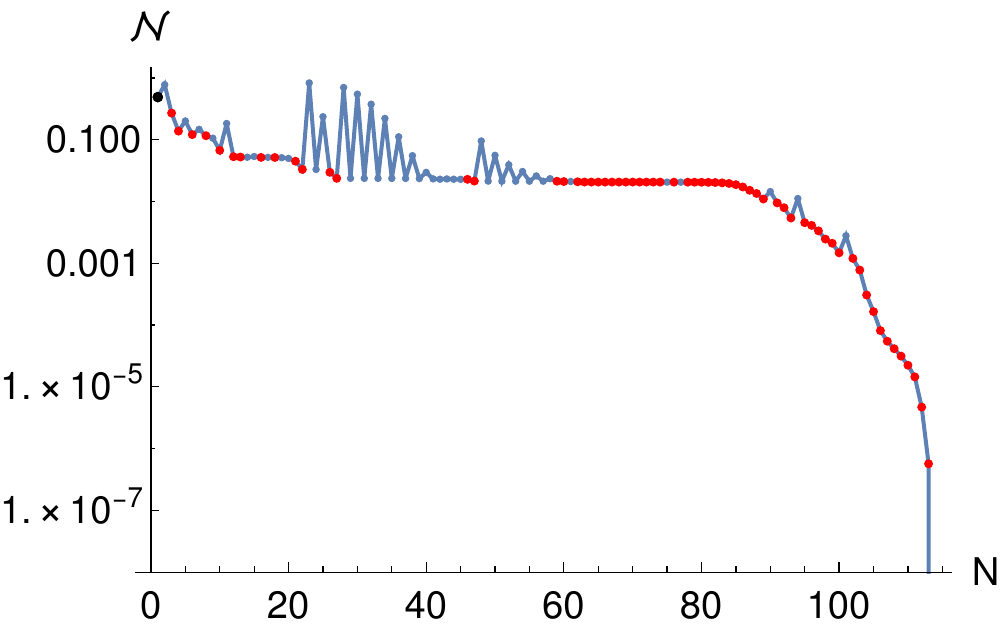}
\caption{} \label{fig:a}
\end{subfigure}\hspace*{\fill}
\begin{subfigure}{0.48\textwidth}
\includegraphics[width=\linewidth]{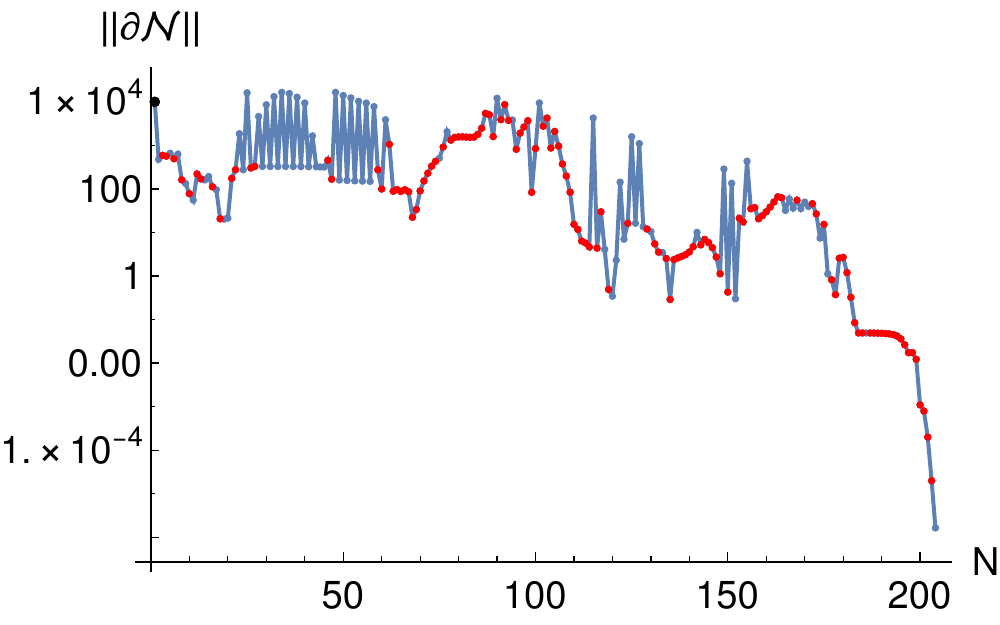}
\caption{} \label{fig:b}
\end{subfigure}

\medskip
\begin{subfigure}{0.48\textwidth}
\includegraphics[width=\linewidth]{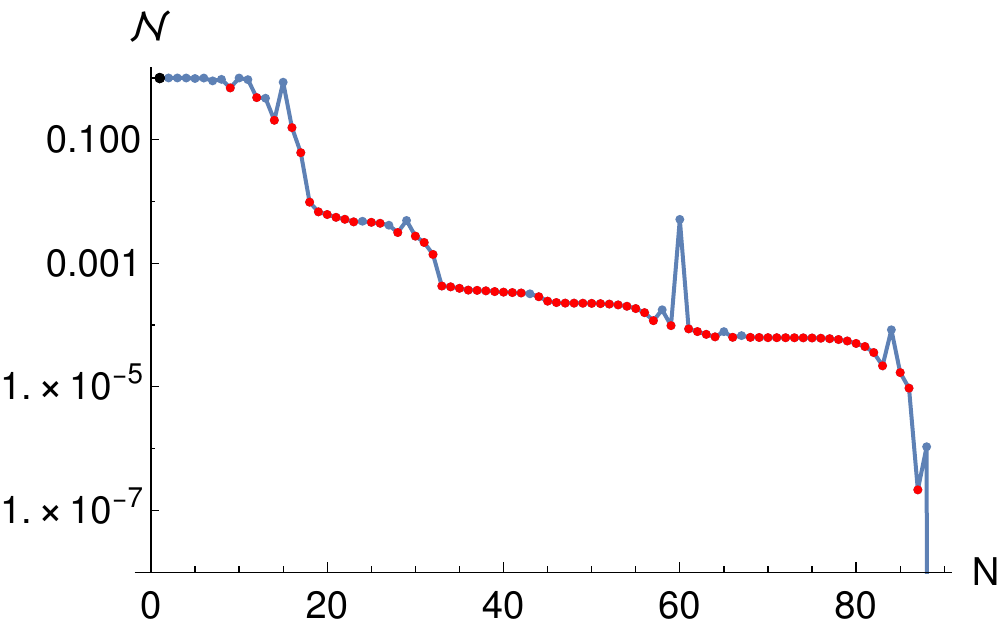}
\caption{} \label{fig:c}
\end{subfigure}\hspace*{\fill}
\begin{subfigure}{0.48\textwidth}
\includegraphics[width=\linewidth]{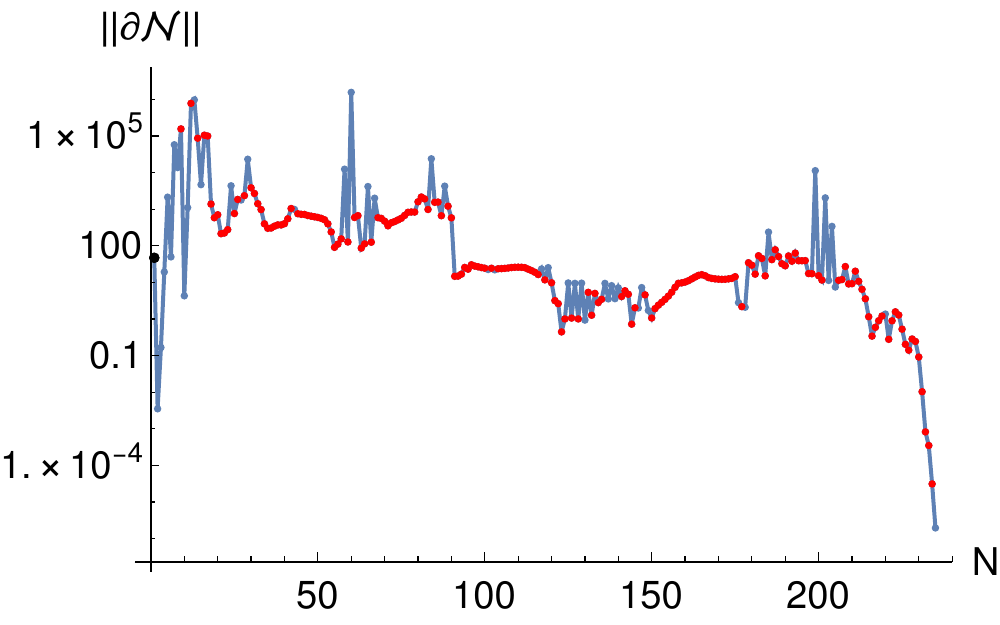}
\caption{} \label{fig:d}
\end{subfigure}

\medskip
\begin{subfigure}{0.48\textwidth}
\includegraphics[width=\linewidth]{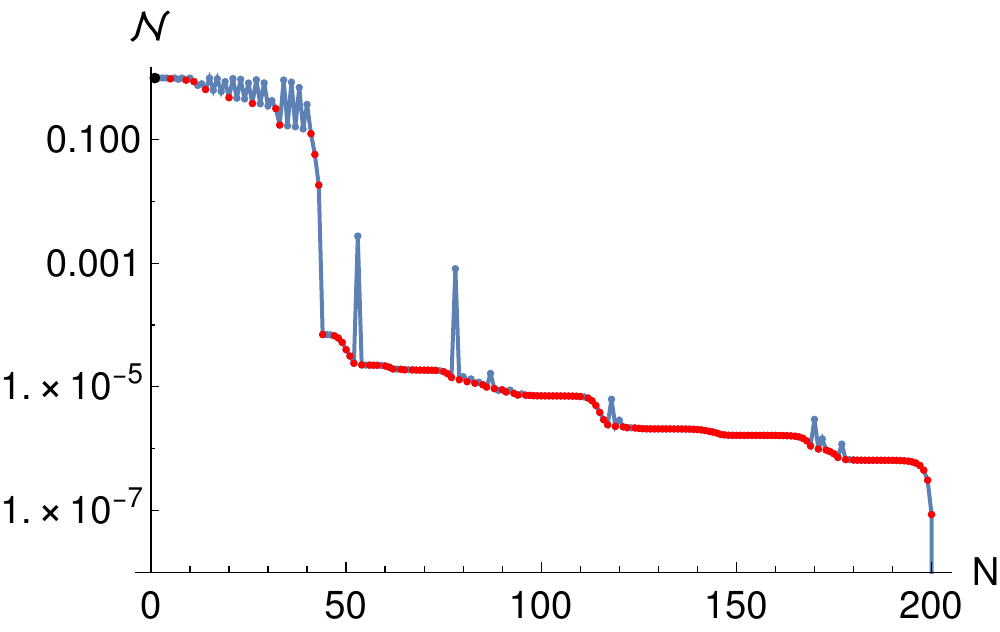}
\caption{} \label{fig:e}
\end{subfigure}\hspace*{\fill}
\begin{subfigure}{0.48\textwidth}
\includegraphics[width=\linewidth]{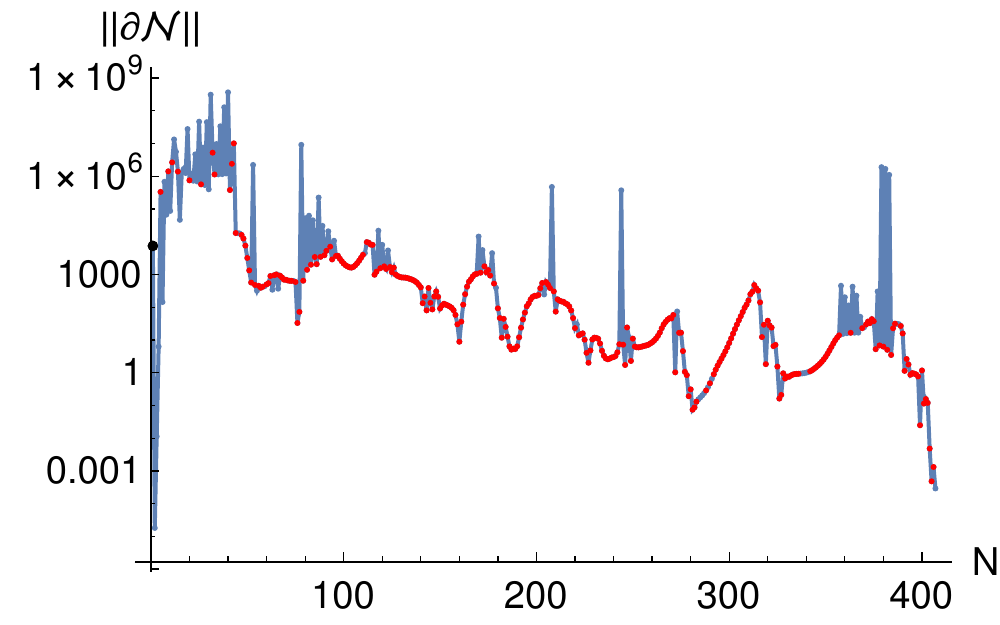}
\caption{} \label{fig:f}
\end{subfigure}

\caption{ (a), (c), (e): Logarithmic plot of the navigator value $\N_i$ at the $i$-th function call in a typical navigator minimization run for respectively a $4$, $7$ and $13$ dimensional parameter space. Only function calls before reaching the negative navigator region are shown. Red dots represent BFGS steps while blue dots indicate function calls within the line searches. (b), (d), (f): Logarithmic plot of the norm of the gradient $\norm{\partial \N}$ at the $i$-th function call for the same navigator searches.} \label{fig:Convergence}
\end{figure}

We can also look in more detail how the navigator searches converge for different values of $N$. In Figures \ref{fig:a}, \ref{fig:c} and \ref{fig:e} we plot the value of the navigator function at the $i$-th function call for a navigator minimization run in parameter spaces of respectively dimensionality $4$, $7$ and $13$. We show the norm of the gradient along these same runs in Figures~\ref{fig:b}, \ref{fig:d} and \ref{fig:f}. You can see that the norm of the gradient falls and rises multiple times along the search. This strongly hints that the navigator function is notably non-convex even in the vicinity of the minimum.  This behaviour gets worse for larger parameter spaces. 

\begin{figure}[htpb] 
\begin{subfigure}{0.48\textwidth}
\includegraphics[width=\linewidth]{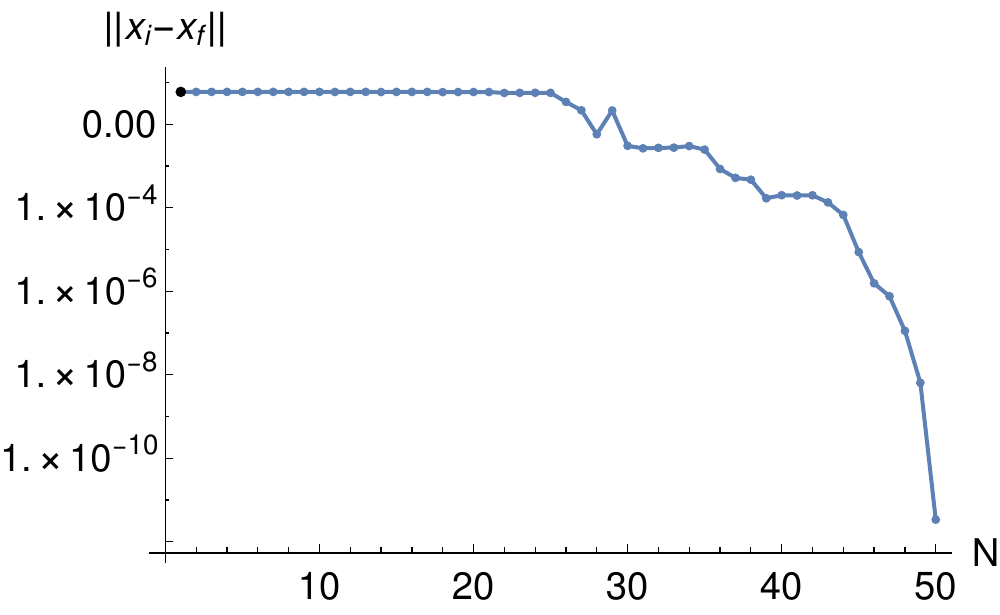}
\caption{} \label{fig:ext_a}
\end{subfigure}\hspace*{\fill}
\begin{subfigure}{0.48\textwidth}
\includegraphics[width=\linewidth]{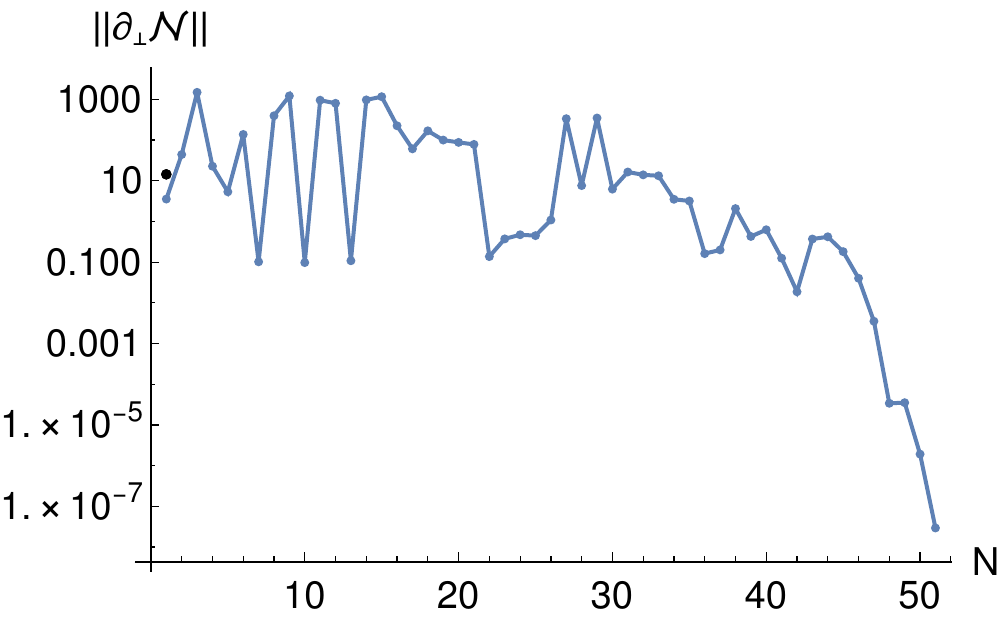}
\caption{} \label{fig:ext_b}
\end{subfigure}

\medskip
\begin{subfigure}{0.48\textwidth}
\includegraphics[width=\linewidth]{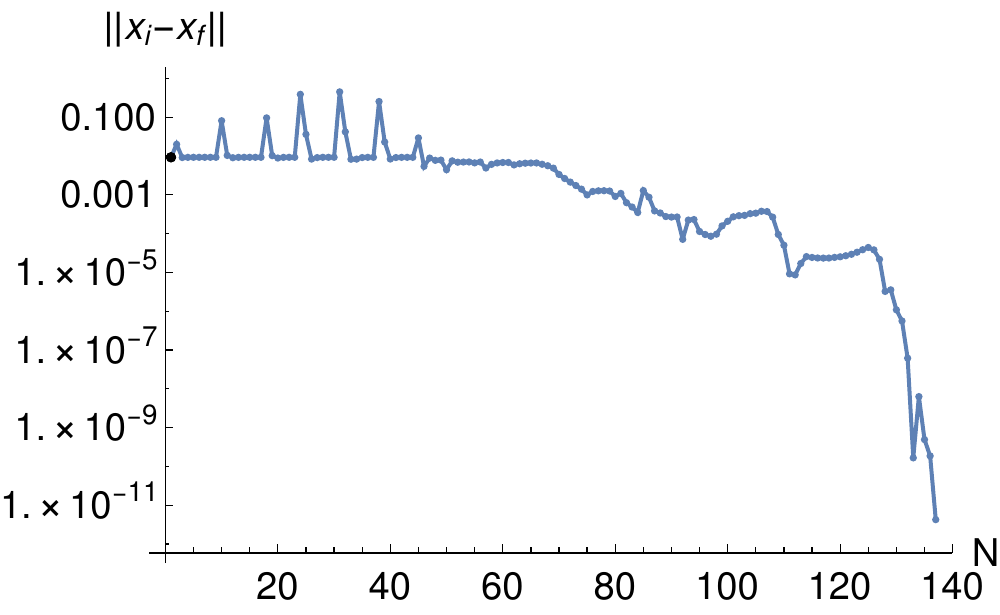}
\caption{} \label{fig:ext_c}
\end{subfigure}\hspace*{\fill}
\begin{subfigure}{0.48\textwidth}
\includegraphics[width=\linewidth]{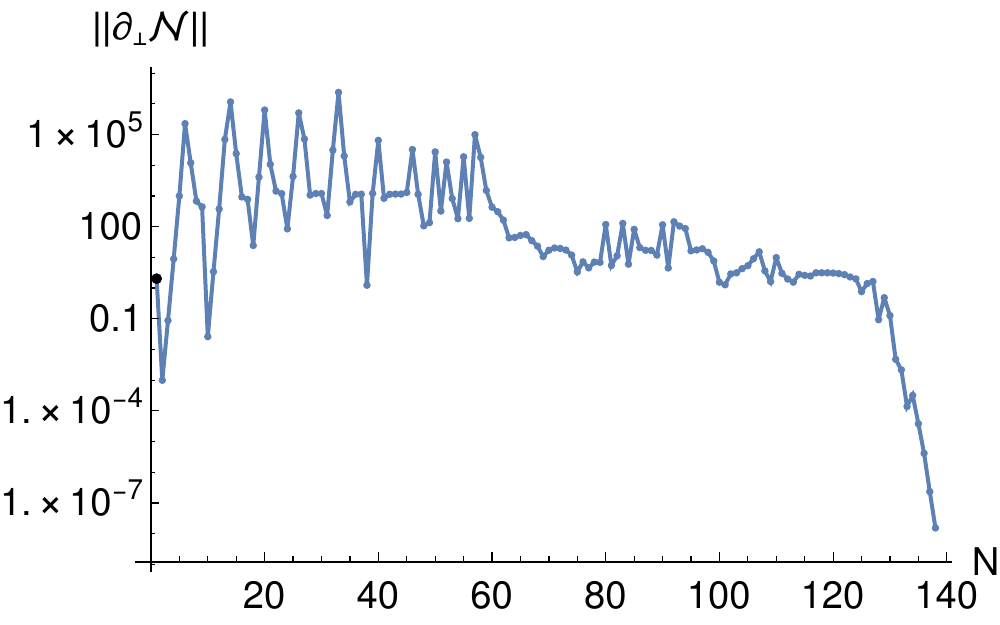}
\caption{} \label{fig:ext_d}
\end{subfigure}

\medskip
\begin{subfigure}{0.48\textwidth}
\includegraphics[width=\linewidth]{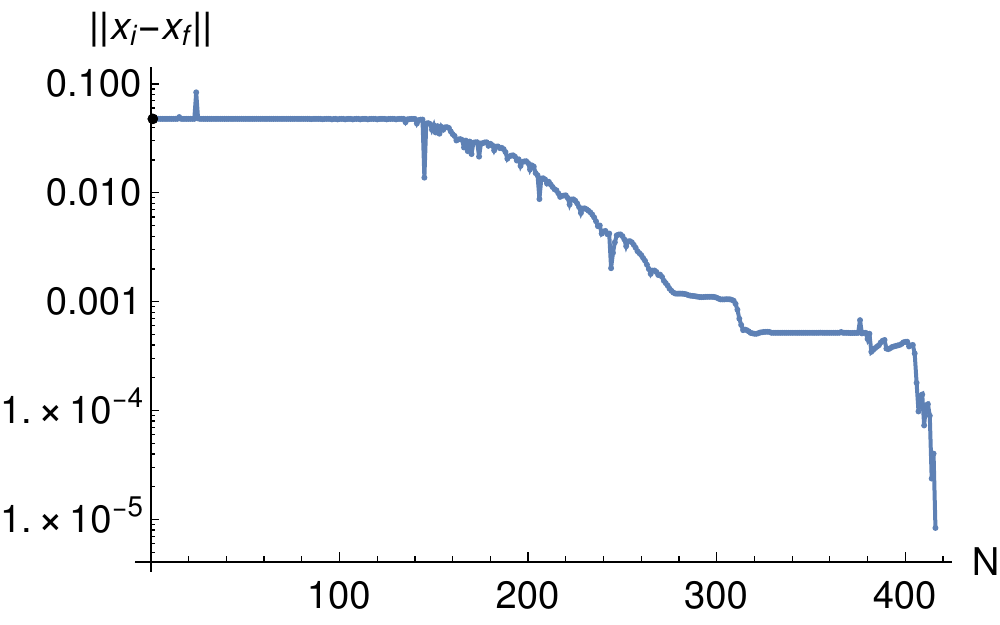}
\caption{} \label{fig:ext_e}
\end{subfigure}\hspace*{\fill}
\begin{subfigure}{0.48\textwidth}
\includegraphics[width=\linewidth]{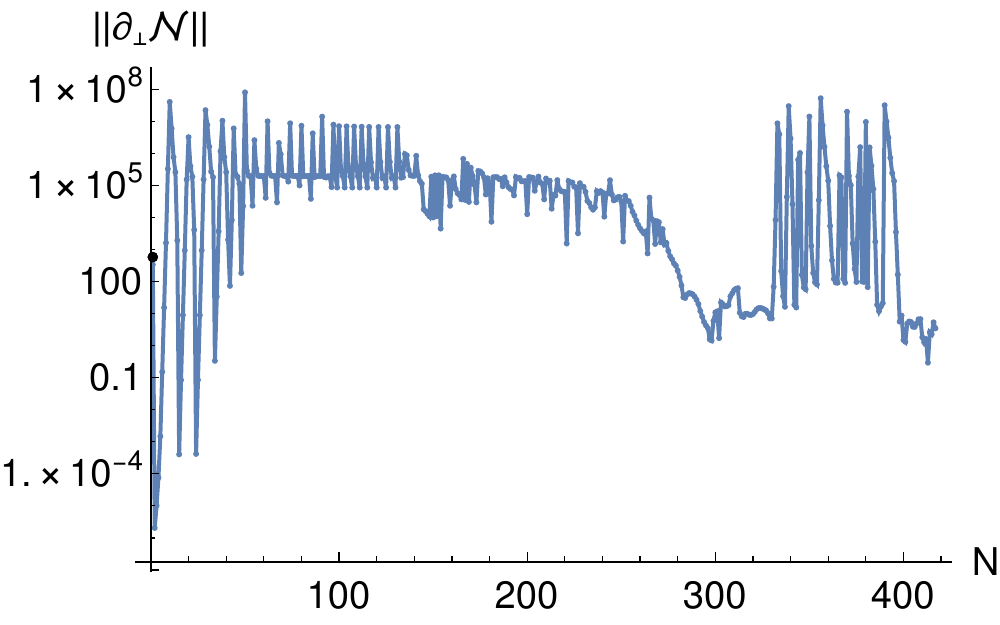}
\caption{} \label{fig:ext_f}
\end{subfigure}

\caption{ (a), (c), (e): Logarithmic plot of the distance $\norm{x_i-x_f}$ between the point $x_i$ and the eventually found extremal point $x_f$ for a typical parameter extremization run in respectively a $4$, $7$ and $13$ dimensional parameter space. Red dots represent BFGS steps while blue dots indicate function calls within the line searches. (b), (d), (f): Logarithmic plot of the norm of the gradient components orthogonal to the direction that is being extremized for the same runs. (The large jumps around point 330 to 390 in the last plot are due to the fact that we reset the approximated Hessian and changed the finite step size used to compute the gradient in the hope of improving convergence.)} \label{}
\end{figure}

For parameter minimization/maximization runs we show in
In Figures~\ref{fig:ext_a}, \ref{fig:ext_c} and \ref{fig:ext_e} we show the distance $\norm{x_i-x_f}$ between the point $x_i$ and the eventually found extremal point $x_f$ for three parameter minimization/maximization runs in parameter spaces of respectively dimensionality $4$, $7$ and $13$. In Figures~\ref{fig:ext_b}, \ref{fig:ext_d} and \ref{fig:ext_f} we show the norm of the gradient components orthogonal to the direction that is being minimized or maximized for those same runs.

\newpage

\section{Conclusions}
\label{sec:conclusions}
We find that the navigator function can efficiently search islands in high-dimensional parameter spaces and obtain rigorous bounds on a large number of parameters. This  opens up many new avenues of exploration and raises the hope that the numerical bootstrap can not just solve for the dimensions and OPEs of the lowest dimensional operators but for any arbitrary operator that is of interest.

We investigated the effect of new sparseness conditions that allow for an explicit scan over many operator dimensions. For many years sparseness of the operator spectrum has been used as a guide to find interesting CFTs. Often the imposed sparseness conditions were physically motivated by experiments revealing the number of relevant operators. Other times the existence of one relevant operator of a certain type was simply assumed because perturbative computations made it plausible and it helped isolate a unique interesting CFT candidate (see for example \cite{Kos:2015mba}\footnote{The assumption that there is only one relevant vector made in that paper, could in principle be tested physically in an experiment where $O(N)$ symmetry emerges and isn't microscopically present.}).  In other cases sparseness was imposed more implicitly, by maximizing the allowed gap on the lowest dimensional operator of a certain type in hopes of identifying kinks corresponding to interesting CFTs.  

Since sparseness of the low dimensional operators is one of the leading principles in the identification and isolation of CFTs in the numerical conformal bootstrap it is important to know the effect of additional and stronger sparseness conditions on numerical conformal bootstrap bounds. The sparseness assumptions that we studied in this paper result in a bootstrap problem that depends on many variables. Previously it would have been impossible to search such a high-dimensional parameter space. Now, the navigator function enables us to do so. These assumptions allow us to find new rigorous bounds on OPE data for which previously there were only non-rigorous estimates (see Table~\ref{table:summary_all_strongest_bounds} for the bounds we obtained). 

In addition searching such high-dimensional spaces could also help us identify new CFT's (by searching for kinks or peaks in high-dimensional search spaces) or help obtain islands where the bootstrap previously was not constraining enough.  

Note also that imposing sparseness conditions on the lowest dimensional operators in certain sectors can help abate the fake primary effect \cite{Karateev:2019pvw} by disconnecting positivity constraints on the various linked sectors.

Similarly sparseness conditions of the form studied in this paper could be very helpful to deal with so called spurious operators which appear in extremal spectra but are not really present in the CFT \cite{Simmons-Duffin:2016wlq}. These operators cause further inaccuracies when solving (truncated) crossing equations for the OPE coefficients of the operators. These spurious operators can be identified as such because their position depends heavily on the exact gap assumptions that were used as well as on the derivative order $\Lambda$. The appearance of spurious operators can be avoided by demanding that there are only $n$ operators below a certain gap. This should also help avoid the related \emph{sharing effect} \cite{Liu:2020tpf} where contributions at the imposed gap cause inaccuracies. 

By pushing the gap to higher dimensions (and scanning over the operators underneath this gap) this contribution should shrink. Moreover, there is more freedom to pick this higher gap to be far away from any physical operator to further diminish the sharing effect.



For all navigator functions we studied we found a unique allowed island exactly around the estimates obtained using the extremal functional method (in addition to a large allowed peninsula at large values of the external dimensions). Moreover, the rigorous error bars seem to be close to the non-rigorous estimate of the "error bars". Thus, these results validate the extremal functional method. This also means that \cite{Simmons-Duffin:2016wlq} correctly identified which operators were "spurious", and can be left out while still satisfying crossing and which are essential. It is a highly non-trivial check that we can still find a (unique) allowed island under assumptions that exclude these spurious operators. 

Surprisingly, one of the rigorous bounds excludes values for $\lambda_{\e\e\e'}$ obtained using \texttt{spectrum} at higher derivative order. We argued that this must be due to the aforementioned sharing effect and that our rigorous bounds are correct.

We believe that we found the minimal and maximal allowed parameter values for all navigator functions and thus found rigorous bounds on these quantities. The value of the navigator and its gradient at the minimal/maximal parameter values provide strong evidence that we at least found the local extremum. The fact that repeated searches from different initial points within the island all converge to the same extremal points makes us believe we did not miss some other extremum at smaller/larger parameter values. This result could be made more rigorous by applying global optimization algorithm that classify all convex regions and find all minima within those regions given a sufficiently dense sampling, but this would come at a significant additional computational cost.

This work shows that navigator search methods can be used as an efficient more rigorous alternative to EFM and that it can be used to provide rigorous error bars on estimates that have already been obtained by EFM. The navigator's ability to verify approximate solutions and provide error bars is of course not limited to the EFM and can be applied to obtain rigorous error bars for any estimate without them. This could become more and more important with the advent of many promising non-rigorous methods of obtaining approximate CFT spectra from truncated crossing equations \cite{Gliozzi:2013ysa,Gliozzi:2014jsa,Gliozzi:2015qsa,Gliozzi:2016cmg,Esterlis:2016psv,Hikami:2017hwv,Hikami:2017sbg,Hikami:2018mrf,Li:2017agi,Li:2017ukc,Leclair:2018trn,El-Showk:2016mxr,PaulosUnpublished,Kantor:2021jpz,Afkhami-Jeddi:2021iuw}.

In \cite{Reehorst:2021ykw} it was only shown that the navigator function offered an efficient method for finding allowed islands and their boundaries in search spaces of dimensionality 2 and 3. The fact that we could find allowed islands and their boundaries in up to 13-dimensional parameter spaces shows that navigator searches scale much better with the dimensionality of search spaces than feasibility based methods and that the numerical conformal bootstrap is no longer tied to low dimensional searches.

\section*{Acknowledgements}  
MR was supported by Mitsubishi Heavy Industries (MHI-ENS Chair) for part of this work and is currently supported by the Simons Foundation grant \#488659 (Simons collaboration on the non-perturbative bootstrap). MR thanks Johan Henriksson, Balt van Rees, Slava Rychkov and David Simmons-Duffin for useful discussions and comments on the draft of this paper.

The computations in this work were performed on the Caltech High Performance Cluster, partially supported by a grant from the Gordon and Betty Moore Foundation.

\appendix    

\section{Navigators}
\label{app:navigators}
Here we give a more detailed description of how to compute the various navigators used in the main text. When something such as $S_-$ is left unspecified for a certain navigator function it means it takes the "default" values as described in section 2.2. of \cite{Reehorst:2021ykw}. Below we write only the GFF terms required for $(\Delta_\sigma,\Delta_\epsilon)$ values in the vicinity of the Ising model. In practice we instead used the more general definition of $\vec{M}_{\rm GFF}$ that includes all GFF operators excluded by the gap assumptions at those specific values of the external dimensions $(\Delta_\sigma,\Delta_\epsilon)$. Close to the Ising model these definitions of course coincide.

\subsection{$\N(\Delta_{\sigma},\Delta_{\epsilon},\theta)$}
This is the same navigator as the one studied in section 2.2 of \cite{Reehorst:2021ykw} except we replaced $\Delta_{\epsilon'}>3$ with $\Delta_{\epsilon'}>6$. This means we have to modify the scalar sector of $S_+$ as well as the GFF navigator:
\begin{equation}
\label{eq:app:nav1}
 \begin{split}
    S_+ \quad &= \{(\Delta,0): \Delta\ge 6\}\cup \{(\Delta,\ell): \ell =2,4,6,\ldots \text{ and }\Delta\ge \ell+1\}\\
    \vec{M}_{\rm GFF} &= \text{Tr}\left[ \begin{pmatrix} 2&0\\0&0 \end{pmatrix} \vec{V}_{+,2\Delta_\s,0}\right]+\text{Tr}\left[ \begin{pmatrix} 0&0\\0&2 \end{pmatrix} \vec{V}_{+,2\Delta_\e,0}\right] + \vec{V}_{-,\Delta_\s+\Delta_\e,0}\, \\
        &+\text{Tr}\left[ \begin{pmatrix} \lambda^2_{\sigma\sigma(2\Delta_\sigma+2,0)}&0\\0&0 \end{pmatrix} \vec{V}_{+,2\Delta_\s+2,0}\right]+\text{Tr}\left[ \begin{pmatrix} 0&0\\0&\lambda^2_{\epsilon\epsilon(2\Delta_\epsilon+2,0)} \end{pmatrix} \vec{V}_{+,2\Delta_\e+2,0}\right] \\
        &+\text{Tr}\left[ \begin{pmatrix} \lambda^2_{\sigma\sigma(2\Delta_\sigma+4,0)}&0\\0&0 \end{pmatrix} \vec{V}_{+,2\Delta_\s+4,0}\right] \, .
 \end{split}
\end{equation}
As explained in \cite{Reehorst:2021ykw} the GFF-navigator is then found by optimizing the following polynomial matrix problem\footnote{The polynomial matrix problem is in turn written as a Semi-Definite Program in the standard manner and solved using SDPB \cite{Simmons-Duffin:2015qma,Landry:2019qug}. See Appendix \ref{app:parameters_numerics} for the choice of numerical parameters we used.} 
\begin{align}
&\N(\Delta_{\sigma},\Delta_{\epsilon},\theta) = \max\ \vec\alpha\cdot \vec{V}_{0,0} \text{ over all linear functionals $\vec \alpha$ such that}\nonumber\\
&\qquad\vec\alpha\cdot \vec{M} = - 1 \,,\nonumber
\\
&\qquad \vec\alpha \cdot \left(
    \vec{V}_{+,\Delta_\e,0}
    +\begin{pmatrix}1&0\\ 0 & 0\end{pmatrix}
    \vec{V}_{-,\Delta_\sigma,0}
    \right)  \succcurlyeq 0\,,\label{torepl}
\\
&\qquad\vec\alpha\cdot \vec{V}_{+,\De,\ell} \succcurlyeq 0
 \text{ for all }(\Delta,\ell)\in S_+\,,\nonumber
 \\
&\qquad\vec\alpha\cdot \vec{V}_{-,\De,\ell} \geq 0
 \text{ for all }(\Delta,\ell)\in S_-\,.\nonumber
\end{align}

\subsection{$\N(\Delta_{\sigma},\Delta_{\epsilon},\theta,\Delta_{\epsilon'})$}
This navigator is the same as the one above in equation \eqref{eq:app:nav1}  except we add $\Delta_{\epsilon'}$ to $S_+$:
\begin{equation}
\label{eq:app:nav2}
      S_+ \quad = \{(\Delta_{\epsilon'},0)\} \cup \{(\Delta,0): \Delta\ge 6\}\cup \{(\Delta,\ell): \ell =2,4,6,\ldots \text{ and }\Delta\ge \ell+1\} \, .
\end{equation}

\subsection{$\N(\Delta_\epsilon,\Delta_\sigma,\lambda_{\sigma \sigma \epsilon},\lambda_{\epsilon \epsilon \epsilon},\Delta_{\epsilon'},\lambda_{\sigma \sigma \epsilon'},\lambda_{\epsilon \epsilon \epsilon'})$}
For this navigator function we contract both the $\epsilon$ and $\epsilon'$ contributions with the respective test values $(\lambda_{\sigma \sigma \epsilon},\lambda_{\epsilon \epsilon \epsilon})$ and $(\lambda_{\sigma \sigma \epsilon'},\lambda_{\epsilon \epsilon \epsilon'})$ adding these contributions with fixed OPE coefficient values to the identity contribution:
\begin{equation}
\begin{split}
    \vec{V}_\textrm{max} =& \vec{V}_{0,0} + \vec{V}_\textrm{fixed} \quad \text{with} \quad,\\
        \vec{V}_\textrm{fixed} =& \text{Tr}\left[ \begin{pmatrix} \lambda_{\sigma \sigma \epsilon}^2 & \lambda_{\sigma \sigma \epsilon} \lambda_{\epsilon \epsilon \epsilon} \\ \lambda_{\sigma \sigma \epsilon} \lambda_{\epsilon \epsilon \epsilon}  & \lambda_{\epsilon \epsilon \epsilon}^2\end{pmatrix} \left(
    \vec{V}_{+,\Delta_\e,0}
    +\begin{pmatrix}1&0\\ 0 & 0\end{pmatrix}
    \vec{V}_{-,\Delta_\sigma,0}
     \right) \right] + \\
     &\qquad \quad + \text{Tr}\left[ \begin{pmatrix} \lambda_{\sigma \sigma \epsilon'}^2 & \lambda_{\sigma \sigma \epsilon'} \lambda_{\epsilon \epsilon \epsilon'} \\ \lambda_{\sigma \sigma \epsilon'} \lambda_{\epsilon \epsilon \epsilon'}  & \lambda_{\epsilon \epsilon \epsilon'}^2\end{pmatrix} \vec{V}_{+,\Delta_{\e'},0}
     \right] \,.
\end{split}
\end{equation}
To guarantee the existance of the GFF-solution to crossing we need to subtract $\vec{V}_\textrm{fixed}$ from  $\vec{M}_{\rm GFF}$
\begin{equation}
    \begin{split}
        \vec{M}_{\rm GFF} &= \text{Tr}\left[ \begin{pmatrix} 2&0\\0&0 \end{pmatrix} \vec{V}_{+,2\Delta_\s,0}\right]+\text{Tr}\left[ \begin{pmatrix} 0&0\\0&2 \end{pmatrix} \vec{V}_{+,2\Delta_\e,0}\right] + \vec{V}_{-,\Delta_\s+\Delta_\e,0}\, \\
        &+\text{Tr}\left[ \begin{pmatrix} \lambda^2_{\sigma\sigma(2\Delta_\sigma+2,0)}&0\\0&0 \end{pmatrix} \vec{V}_{+,2\Delta_\s+2,0}\right]+\text{Tr}\left[ \begin{pmatrix} 0&0\\0&\lambda^2_{\epsilon\epsilon(2\Delta_\epsilon+2,0)} \end{pmatrix} \vec{V}_{+,2\Delta_\e+2,0}\right] \\
        &+\text{Tr}\left[ \begin{pmatrix} \lambda^2_{\sigma\sigma(2\Delta_\sigma+4,0)}&0\\0&0 \end{pmatrix} \vec{V}_{+,2\Delta_\s+4,0}\right] - \vec{V}_\textrm{fixed} \,.
    \end{split}
\end{equation}
Now we maximize this $\vec{V}_\textrm{max}$, removing the $\Delta_{\epsilon}$ and $\Delta_{\epsilon'}$ contributions from the other positivity conditions since they are already accounted for here. This leads to the following polynomial matrix problem:
\begin{align}
 & S_+ \quad = \{(\Delta,0): \Delta\ge 6\}\cup \{(\Delta,\ell): \ell =2,4,6,\ldots \text{ and }\Delta\ge \ell+1\}\\
&\N(\Delta_\epsilon,\Delta_\sigma,\lambda_{\sigma \sigma \epsilon},\lambda_{\epsilon \epsilon \epsilon},\Delta_{\epsilon'},\lambda_{\sigma \sigma \epsilon'},\lambda_{\epsilon \epsilon \epsilon'}) = \max\ \vec\alpha\cdot \vec{V}_{\textrm{max}} \text{ over all linear functionals $\vec \alpha$ such that}\nonumber\\
&\qquad\vec\alpha\cdot \vec{M} = - 1 \,,\nonumber
\\
&\qquad\vec\alpha\cdot \vec{V}_{+,\De,\ell} \succcurlyeq 0
 \text{ for all }(\Delta,\ell)\in S_+\,,\nonumber
 \\
&\qquad\vec\alpha\cdot \vec{V}_{-,\De,\ell} \geq 0
 \text{ for all }(\Delta,\ell)\in S_-\,.\nonumber
\end{align}

\subsection{$\N(\Delta_\epsilon,\Delta_\sigma,\theta,\Delta_{\sigma'})$}
Here we instead assume $\Delta_{\epsilon'}>6$ and $\Delta_{\epsilon'}>3$. Thus, compared to the standard setup studied in \cite{Reehorst:2021ykw} we have to modify $S_-$ as well as $\vec{M}_{\rm GFF}$:
\begin{equation}
\label{eq:app:navo1}
 \begin{split}
    S_- \quad &= \{(\Delta_{\sigma'},0)\} \cup \{(\Delta,0): \Delta\ge 6\}\cup \{(\Delta,\ell): \ell =2,4,6,\ldots \text{ and }\Delta\ge \ell+1\}\\
    \vec{M}_{\rm GFF} &= \text{Tr}\left[ \begin{pmatrix} 2&0\\0&0 \end{pmatrix} \vec{V}_{+,2\Delta_\s,0}\right]+\text{Tr}\left[ \begin{pmatrix} 0&0\\0&2 \end{pmatrix} \vec{V}_{+,2\Delta_\e,0}\right] + \vec{V}_{-,\Delta_\s+\Delta_\e,0}\, \\
        & + \lambda^2_{\sigma\epsilon(\Delta_\s+\Delta_\e+2,0)} \vec{V}_{-,\Delta_\s+\Delta_\e+2,0} + \lambda^2_{\sigma\epsilon(\Delta_\s+\Delta_\e+4,0)} \vec{V}_{-,\Delta_\s+\Delta_\e+4,0} \,.
 \end{split}
\end{equation}

\subsection{$\N(\Delta_\epsilon,\Delta_\sigma,\theta,\Delta_{\sigma'},\lambda_{\sigma\epsilon \sigma'})$}
We can scan over the OPE magnitude $\lambda^2_{\sigma\epsilon \sigma'}$ by adding its fixed contribution to $\vec{V}_\textrm{max}$ (removing it from $S_-$ and subtracting the same contribution from $\vec{M}_{\rm GFF}$ to guarantee boundedness):
\begin{equation}
\begin{split}
        S_- \hspace{14pt} &= \{(\Delta_{\sigma'},0)\} \cup \{(\Delta,0): \Delta\ge 6\}\cup \{(\Delta,\ell): \ell =2,4,6,\ldots \text{ and }\Delta\ge \ell+1\}\\
        \vec{V}_\textrm{max} \hspace{6pt} &= \vec{V}_{0,0} +  \lambda^2_{\sigma\epsilon \sigma'} \vec{V}_{-,\Delta_{\sigma'},0} \, , \\
        \vec{M}_{\rm GFF} &= \text{Tr}\left[ \begin{pmatrix} 2&0\\0&0 \end{pmatrix} \vec{V}_{+,2\Delta_\s,0}\right]+\text{Tr}\left[ \begin{pmatrix} 0&0\\0&2 \end{pmatrix} \vec{V}_{+,2\Delta_\e,0}\right] + \vec{V}_{-,\Delta_\s+\Delta_\e,0}  \\
        &+ \lambda^2_{\sigma\epsilon(\Delta_\s+\Delta_\e+2,0)} \vec{V}_{-,\Delta_\s+\Delta_\e+2,0} + \lambda^2_{\sigma\epsilon(\Delta_\s+\Delta_\e+4,0)} \vec{V}_{-,\Delta_\s+\Delta_\e+4,0} - \lambda^2_{\sigma\epsilon \sigma'} \vec{V}_{-,\Delta_{\sigma'},0} \,.
\end{split}
\end{equation}

\subsection{$\N(\Delta_\epsilon,\Delta_\sigma,\theta,c_T)$}
This is the case described in equation \eqref{eq:crossing_ct}. We make the additional assumption $\Delta_{T'}>4$. Thus, we have to modify $S_+$ as and include the excluded spin-2 operators in $\vec{M}_{\rm GFF}$. We also add the fixed contribution  $\vec{V}_\textrm{fixed}= \tilde{p}_T \tilde{V}_{+,3,2} $ to the identity (see the paragraph below equation \eqref{eq:fixed_term}) and therefore we also have to subtract it from $\vec{M}_{\rm GFF}$. This gives the following polynomial matrix problem.
\begin{equation}
    \begin{split}
              S_+ \quad &=\{(\Delta,0): \Delta\ge 3\}\cup  \{(\Delta,2) : \Delta>4\} \cup \{(\Delta,\ell): \ell =4,6,\ldots \text{ and }\Delta\ge \ell+1\} \, \\          
              \vec{M}_{\rm GFF} &= \text{Tr}\left[ \begin{pmatrix} 2&0\\0&0 \end{pmatrix} \vec{V}_{+,2\Delta_\s,0}\right]+\text{Tr}\left[ \begin{pmatrix} 0&0\\0&2 \end{pmatrix} \vec{V}_{+,2\Delta_\e,0}\right] + \vec{V}_{-,\Delta_\s+\Delta_\e,0}\, \\
        &+ \text{Tr}\left[ \begin{pmatrix} \lambda^2_{\sigma\sigma(2\Delta_\sigma+2,2)}&0\\0&0 \end{pmatrix} \vec{V}_{+,2\Delta_\s+2,2}\right] - \vec{V}_\textrm{fixed} \,.
    \end{split}
\end{equation}

\subsection{$\N(\Delta_\epsilon,\Delta_\sigma,\theta,c_T, \Delta_{T'})$}
Here we instead assume $\Delta_{T''}>6$. Thus, we need to add extra terms to $\vec{M}_{\rm GFF}$ and modify $S_+$ again. We also have to add the $T'$ contribution to $\vec{V}_\textrm{fixed}$.
\begin{equation}
    \begin{split}
              S_+ \quad &=\{(\Delta,0): \Delta\ge 3\}\cup \{(\Delta,2) : \Delta>6\} \cup  \{(\Delta,\ell): \ell =4,6,\ldots \text{ and }\Delta\ge \ell+1\} \, \\     
    \vec{V}_\textrm{fixed} &= \tilde{p}_T \tilde{V}_{+,3,2}+ \text{Tr}\left[ \begin{pmatrix} \lambda_{\sigma \sigma T'}^2 &         \lambda_{\sigma \sigma T'} \lambda_{\epsilon \epsilon T'} \\ \lambda_{\sigma         \sigma T'} \lambda_{\epsilon \epsilon T'}  & \lambda_{\epsilon \epsilon               T'}^2\end{pmatrix} 
         \vec{V}_{+,\Delta_{T'},2} \right]  \\
             \vec{M}_{\rm GFF} &= \text{Tr}\left[ \begin{pmatrix} 2&0\\0&0 \end{pmatrix} \vec{V}_{+,2\Delta_\s,0}\right]+\text{Tr}\left[ \begin{pmatrix} 0&0\\0&2 \end{pmatrix} \vec{V}_{+,2\Delta_\e,0}\right] + \vec{V}_{-,\Delta_\s+\Delta_\e,0}\, \\
        &+ \text{Tr}\left[ \begin{pmatrix} \lambda^2_{\sigma\sigma(2\Delta_\sigma+2,2)}&0\\0&0 \end{pmatrix} \vec{V}_{+,2\Delta_\s+2,2}\right] + 
        \text{Tr}\left[ \begin{pmatrix} 0&0\\0&\lambda^2_{\epsilon\epsilon(2\Delta_\epsilon+2,2)} \end{pmatrix}                      \vec{V}_{+,2\Delta_\e+2,2}\right] \\
        &+ \text{Tr}\left[ \begin{pmatrix} \lambda^2_{\sigma\sigma(2\Delta_\sigma+4,2)}&0\\0&0 \end{pmatrix} \vec{V}_{+,2\Delta_\s+4,2}\right] - \vec{V}_\textrm{fixed}  \,.
    \end{split}
\end{equation}

\subsection{$\N(\Delta_\epsilon,\Delta_\sigma,\lambda_{\sigma \sigma \epsilon},\lambda_{\epsilon \epsilon \epsilon},\Delta_{\epsilon'},\lambda_{\sigma \sigma \epsilon'},\lambda_{\epsilon \epsilon \epsilon'},\Delta_{\sigma'},\lambda_{\sigma\epsilon \sigma'},c_T, \Delta_{T'},\lambda_{\sigma\sigma T'},\lambda_{\epsilon\epsilon T'})$}
Here we put all the gap assumptions together. So we assume there are exactly two operator with a dimension below 6 in the sectors $\bZ_2$-even scalars, $\bZ_2$-odd scalars, and $\bZ_2$-even spin-2 operators. Since we scan over all the OPEs of the isolated operators we add all these contributions to the identity. This gives the following PMP:

\begin{equation}
\begin{split}
    S_+ \hspace{5pt} =&\{(\Delta,0): \Delta\ge 6\}\cup \{(\Delta,2) : \Delta>6\} \cup  \{(\Delta,\ell): \ell =4,6,\ldots \text{ and }\Delta\ge \ell+1\} \, \\   
    S_- \hspace{5pt} =& \{(\Delta,0): \Delta\ge 6\}\cup \{(\Delta,\ell): \ell =2,4,6,\ldots \text{ and }\Delta\ge \ell+1\}\\
    \vec{V}_\textrm{max} =& \vec{V}_{0,0} + \vec{V}_\textrm{fixed} \quad \text{with} \quad,\\
        \vec{V}_\textrm{fixed} =& \text{Tr}\left[ \begin{pmatrix} \lambda_{\sigma \sigma \epsilon}^2 & \lambda_{\sigma \sigma \epsilon} \lambda_{\epsilon \epsilon \epsilon} \\ \lambda_{\sigma \sigma \epsilon} \lambda_{\epsilon \epsilon \epsilon}  & \lambda_{\epsilon \epsilon \epsilon}^2\end{pmatrix} \left(
    \vec{V}_{+,\Delta_\e,0}
    +\begin{pmatrix}1&0\\ 0 & 0\end{pmatrix}
    \vec{V}_{-,\Delta_\sigma,0}
     \right) \right] \\
     &\quad+ \text{Tr}\left[ \begin{pmatrix} \lambda_{\sigma \sigma \epsilon'}^2 & \lambda_{\sigma \sigma \epsilon'} \lambda_{\epsilon \epsilon \epsilon'} \\ \lambda_{\sigma \sigma \epsilon'} \lambda_{\epsilon \epsilon \epsilon'}  & \lambda_{\epsilon \epsilon \epsilon'}^2\end{pmatrix} \vec{V}_{+,\Delta_{\e'},0}
     \right] \, + \lambda^2_{\sigma\epsilon \sigma'} \vec{V}_{-,\Delta_{\sigma'},0}  + \tilde{p}_T \tilde{V}_{+,3,2}  \\
     &\quad + \text{Tr}\left[ \begin{pmatrix} \lambda_{\sigma \sigma T'}^2 &         \lambda_{\sigma \sigma T'} \lambda_{\epsilon \epsilon T'} \\ \lambda_{\sigma         \sigma T'} \lambda_{\epsilon \epsilon T'}  & \lambda_{\epsilon \epsilon               T'}^2\end{pmatrix} 
         \vec{V}_{+,\Delta_{T'},2} \right]
\end{split}
\end{equation}
Moreover,  $\vec{M}_{\rm GFF}$ is in this case constructed by including all terms appearing in the above definitions of $\vec{M}_{\rm GFF}$ (exactly once).

\section{Parameters for numerics}
\label{app:parameters_numerics}

The computation of the navigator function was written as a semidefinite program that was solved using the arbitrary precision solver \texttt{SDPB} \cite{Simmons-Duffin:2015qma,Landry:2019qug}. We  used \texttt{simpleboot} \cite{simpleboot} to set up the SDPs. The parameters used for the computations are presented in Table~\ref{tab:sdpb_parameters}. 
\begin{table}[h]
\begin{center}
\resizebox{0.94\linewidth}{!}{
\begin{tabular}{ |c|c|c|c| } 
\hline
 \texttt{$\Lambda$} & 11 & 19 & 31\\ 
 \texttt{keptPoleOrder}
 & 14 & 15 & 18\\ 
 \texttt{order} & 28 & 30 & 60\\ 
 \texttt{spins} & $\{0,\ldots,27\}$ & $\{0,\ldots,30, 39,40, 49, 50\}$ & $\{0,\ldots,30, 39,40, 49, 50,59,60\}$\\ 
 \texttt{precision} & 896 & 896 & 896   \\ 
 \texttt{dualityGapThreshold} & $10^{-30}$ & $10^{-30}$ & $10^{-30}$ \\ 
 \texttt{primalErrorThreshold} & $10^{-30}$ & $10^{-30}$ & $10^{-30}$\\ 
 \texttt{dualErrorThreshold} & $10^{-30}$ & $10^{-30}$ & $10^{-30}$\\ 
 \texttt{initialMatrixScalePrimal} & $10^{20}$ & $10^{40}$ & $10^{40}$ \\ 
 \texttt{initialMatrixScaleDual} & $10^{20}$ &  $10^{40}$  & $10^{40}$\\ 
 \texttt{feasibleCenteringParameter} & 0.1 & 0.1 & 0.1 \\ 
 \texttt{infeasibleCenteringParameter} & 0.3 & 0.3 & 0.3 \\ 
 \texttt{stepLengthReduction} & 0.7 & 0.7 & 0.7\\ 
 \texttt{maxComplementarity} & $10^{100}$ & $10^{100}$  & $10^{100}$ \\ 
 \hline
\end{tabular}
}
\caption[]{\label{tab:sdpb_parameters}  Parameters used to setup the SDPs, along with the \texttt{SDPB} parameters. The definition of these can be found in \cite{Simmons-Duffin:2015qma} (where \texttt{order} was 90 and \texttt{keptPoleOrder} was $\kappa$).
}
\end{center}
\end{table}

\clearpage
\small
\bibliographystyle{utphys}
\bibliography{navigator}

\end{document}